\documentclass{aa}

\usepackage{graphicx}
\usepackage{txfonts}
\usepackage{natbib}
\bibpunct[]{(}{)}{;}{a}{}{,}

\newcommand{\snia}{SN~Ia}
\newcommand{\req}[1]{Eq.~(\ref{#1})}
\newcommand{\dd}{\ensuremath\mathrm{d}}
\newcommand{\ra}{\ensuremath{\rightarrow}}
\newcommand{\Ra}{\ensuremath{\Rightarrow}}
\newcommand{\pdiff}[2]{\ensuremath{\frac{\partial #1}{\partial #2}}}
\newcommand{\diff}[2]{\ensuremath{\frac{\mathrm{d} #1}{\mathrm{d} #2}}}
\newcommand{\sdiff}[2]{\ensuremath{\frac{\mathrm{d}^{2} #1}{\mathrm{d} #2^{2}}}}
\newcommand{\half}{\ensuremath{{\textstyle\frac{1}{2}}}}
\newcommand{\bti}{\tilde{B}}
\newcommand{\sti}{\tilde{S}}
\newcommand{\jti}{\tilde{J}}
\newcommand{\hti}{\tilde{H}}
\newcommand{\tti}{\tilde{\tau}}
\newcommand{\bnu}{\ensuremath{B_{\nu}}}
\newcommand{\mr}[1]{\ensuremath\mathrm{#1}}
\newcommand{\const}{\ensuremath\mathrm{const}}
\newcommand{\third}{\ensuremath{{\textstyle\frac{1}{3}}}}
\newcommand{\cofs}{\ensuremath{^{56}\mathrm{Co}}}
\newcommand{\nifs}{\ensuremath{^{56}\mathrm{Ni}}}
\newcommand{\fefs}{\ensuremath{^{56}\mathrm{Fe}}}
\newcommand{\msun}{\ensuremath{M_{\odot}}}
\newcommand{\kms}{\ensuremath{\,\mathrm{km}\,\mathrm{s}^{-1}}}
\newcommand{\ergs}{\ensuremath{\,\mathrm{erg}\,\mathrm{s}^{-1}}}
\begin{document}

\title{Non-LTE models for synthetic spectra of type Ia supernovae /\\
hot stars with extremely extended atmospheres}
\subtitle{II. Improved lower boundary conditions for the\\
          numerical solution of the radiative transfer}

\titlerunning{Non-LTE radiative transfer models of SN~Ia}

\author{
D.~N.~Sauer \inst{1,2,3}
\and
T.~L.~Hoffmann \inst{3}
\and
A.~W.~A.~Pauldrach \inst{3}
}

\offprints{D.~Sauer}

\institute{
INAF, Osservatorio Astronomico di Trieste, Via G.~B.~Tiepolo, 11
I-34131 Trieste, Italy\\
\email{sauer@ts.astro.it}
\and
Max-Planck-Institut f\"{u}r Astrophysik,
Karl-Schwarzschild-Str. 1, D-85741 Garching, Germany
\and
Universit\"{a}tssternwarte M\"{u}nchen, Scheinerstr. 1,
D-81679 M\"{u}nchen, Germany\\
\email{hoffmann@usm.uni-muenchen.de{\rm; }uh10107@usm.uni-muenchen.de}
}
\date{Received; accepted }

\abstract
{Realistic atmospheric models that link the properties and the physical
conditions of supernova ejecta to observable spectra are required for the
quantitative interpretation of observational data of type~Ia supernovae (\snia)
and the  assessment of the physical merits of theoretical supernova explosion
models. The numerical treatment of the radiation transport -- yielding the
synthetic spectra -- in models of {\snia} ejecta in early phases is usually
carried out in analogy to atmospheric models of `normal' hot stars. Applying
this analogy indiscriminately leads to inconsistencies in {\snia} models
because a diffusive lower boundary, while justified for hot stars, is invalid
for hydrogen and helium-deficient supernova ejecta. In type~Ia supernovae the
radiation field does not thermalize even at large depths, and large optical
depths are not reached at all wavelengths.}
{We aim to derive an improved description of the lower boundary that allows a
more consistent solution of the radiation transfer in {\snia} and therefore
yields more realistic synthetic spectra.}
{We analyze the conditions that lead to a breakdown of the conventional
diffusion approximation as the lower boundary in {\snia}.  For the radiative
transfer, we use a full non-LTE code originally developed for radiatively
driven winds of hot stars, with adaptations for the physical conditions in
SN~Ia.  In addition to a well-tested treatment of the underlying microphysical
processes, this code allows a direct comparison of the results for {\snia} and
hot stars.}
{We develop a semi-analytical description that allows us to overcome some of
the limiting assumptions in the conventional treatment of the lower boundary in
{\snia} radiative transfer models. We achieve good agreement in a comparison
between the synthetic spectrum of our test model and an observed spectrum. }
{}


\maketitle

\section{Introduction}

Type~Ia supernovae (\snia) have become an invaluable tool for the determination
of the cosmological parameters \citep{riess98, perlmutter99, riess01, tonry03}
as their exceptional brightness makes them observable even at large
cosmological distances.  Using {\snia} for distance determination requires
knowledge of their absolute luminosities; however, {\snia} are not perfect
``standard candles'' in this respect because they show an intrinsic scatter in
their properties, in particular in the peak brightness. The application of
{\snia} for cosmology therefore relies on empirical relationships between the
peak brightness and other observed characteristics \citep[e.g.,][]{phillips93,
hamuy96b, riess96, perlmutter97}.   To first order the shape of the light
curves and certain spectral properties is determined by the mass of synthesized
{\nifs} and its distribution within the ejecta
\citep{nugent95b,hoeflich95,pinto00a}. However, the details of the physical
processes that cause the observed variation are still unclear (see
\citealt{hillebrandt00} for a review).  This uncertainty is an essential
problem because the different calibration methods used to derive the peak
brightness yield partly different results, implying an unaccounted source of
systematic error \citep{leibundgut04}.  In addition, the application of {\snia}
for cosmological distance measurement relies on the crucial assumption that
objects observed at high redshifts have the same properties as objects in the
nearby universe, so that the same calibration method for the luminosity
differences can be applied to all objects.  It is, therefore, of fundamental
interest to derive a physical model that can explain the explosion mechanism of
{\snia} in detail, including the observed intrinsic variability. This will
allow a more reliable estimate of the systematic errors of the distance
measurement.

The currently favored models for the explosion mechanism of {\snia} involve the
thermonuclear explosion of a carbon-oxygen white dwarf (WD) in a binary system.
Two progenitor scenarios are generally considered: the ``single degenerate''
scenario and the ``double degenerate'' scenario. In the single degenerate
scenario the WD accretes  mass from a red-giant companion star. When the WD
reaches a mass close to the Chandrasekhar mass ($M_{\rm Ch}\approx 1.4\msun$)
the compressional heating at the center of the star triggers carbon burning.
After a period of a few thousand years of quiet burning a thermonuclear runaway
disrupts the star \citep{iben84, webbink84, woosley04, han04}. In the ``double
degenerate'' scenario the companion star is also a WD with a mass such that the
total mass of the system exceeds $M_{\rm Ch}$.  Due to energy loss by
gravitational waves the orbital separation of the binary system gradually
decreases, leading to a merger, which triggers the thermonuclear runaway that
explodes the star \citep{whelan73, nomoto82}. Potential progenitor systems have
been found in recent years; their numbers, however, are too low to explain the
observed rates of SN Ia \citep{cappellaro99}. \citet{pauldrach05}  suggests on
basis of the results of \citet{pauldrach04} a potential connection of {\snia}
progenitors to a subgroup of central stars of planetary nebulae (CSPN).

The details of the explosion process itself are also still subject to a lively
debate.  The general picture is that a subsonic deflagration wave (``flame'')
is ignited near the center of the star. The flame travels outward, burning part
of the star to nuclear statistical equilibrium (NSE). Because the flame
propagates subsonically, the star can expand while undergoing burning. This
allows partial burning of C and O to intermediate mass elements (Si, S, Mg, Ca)
which dominate the composition. In contrast to the deflagration, a prompt,
supersonic detonation of the star would generate predominantly iron group
elements. This outcome contradicts the observed composition.  No agreement has
been reached in the debate whether the explosion continues subsonically until
the end of the explosion
\citep{nomoto84,woosley84,niemeyer95,reinecke02,roepke05,roepke05c} or if a
(yet unknown) mechanism triggers the deflagration to turn into a supersonic
detonation toward the end of the explosion (delayed detonation transition, DDT)
\citep{hoeflich96,iwamoto99}. At present, the resulting composition of DDT
models seem to favor the latter scenario; however, those models are not
self-consistent and depend on ad hoc assumptions of the occurrence of the DDT.

Judging the validity of numerical explosion models requires a comparison of the
observational consequences predicted by the explosion models to the {\em de
facto} observations of {\snia}. Realistic radiative transfer models provide the
crucial link between explosion models and observations. It is possible to
predict the observational implications of the hydrodynamical models only if the
radiative transfer models are sufficiently realistic and detailed.  Conversely,
such radiative transfer models make it possible to establish constraints on the
composition and structure of {\snia} from the spectroscopic interpretation of
observed spectra.

{\snia} in phases before and shortly after maximum exhibit a spectrum dominated
by a few very broad absorption features in a non-thermal continuum.  These
absorption features are mostly due to blends of several lines, while the
``continuum'' itself is formed by the overlap of a large number of
Doppler-broadened metal lines. The true continuum opacities and emissivities
that determine the overall shape of the spectrum in stars are of minor
significance in supernovae. At later epochs, the ``pseudo-continuum''
photosphere recedes deeper and deeper into the ejecta and eventually disappears
as the ejecta become transparent. Unlike stars, the radiation in {\snia} is
generated within the expanding medium itself by the deposition of the energy
of $\gamma$-photons that result from the decay chain
{\nifs}\ra{\cofs}\ra{\fefs} \citep{colgate69}.

The early epochs where a photosphere is still present are referred to as the
photospheric phase. In the photospheric phase the ejecta can be treated in
analogy to hot stars as expanding, extended atmospheres. Only the radiative
transfer in the photosphere and in the outer envelope above the photosphere
needs to be considered. In this setup a steady state is assumed because the
photon escape timescale in the thin medium generally will be much shorter than
the expansion timescale.

In principle, a complete radiative transfer model for {\snia} would require
consistent, time-dependent solutions of the populations of all atomic levels
and the continuum and line transfer, including the treatment of energy
deposition by the decay products of {\nifs} and {\cofs}.

Current models implement various simplifications according to the specific
model's purpose.  These simplifications are necessary because the solution of
time-dependent radiative transfer in three dimensions, including the full
coupling of radiation and matter, is not yet feasible and some of the terms
involved in such a consistent solution are shown to be or regarded to be of second
order.  In recent decades synthetic spectra of {\snia} have been modeled by
several groups with a variety of approaches involving different levels of
complexity depending on the application \citep{branch85, mazzali93a,
mazzali93b, eastman93, hoeflich95, nugent95a, pauldrach96, nugent97, lentz01,
hoeflich05, stehle05, baron06,kasen06}.

Highly parametrized models, which implement a simplified treatment of physical
processes to achieve short run-times seem to be suitable for the comparative
analysis of a large number of observed spectra, while more realistic models are
required for a deeper understanding of the physical effects leading to specific
observed properties. In particular, judging the validity of hydrodynamic
explosion models, as mentioned above, can only be performed using radiative
transfer models that include a very detailed treatment of relevant physical
processes.  Such detailed models may also be used to validate or invalidate
specific simplifying assumptions used in less elaborate models.

In spite of the analogy mentioned before, {\snia} ejecta differ from ordinary
stellar atmospheres in several important aspects. Techniques and methods that
are adopted for stellar atmosphere modeling must be carefully checked to verify
whether the applied approximations are still justified for {\snia}. In this
work we present an improved description for the lower boundary of the radiative
transfer calculation.  This is required because the assumption of
thermalization at large depths, justified for stellar atmospheres, breaks down
for {\snia}.

In the work we present here, we use the computer program {\em WMbasic}
\citep{pauldrach94, pauldrach01} to obtain a consistent solution of the full
non-LTE rate equations and a detailed observer's frame solution of the
radiative transfer.  This code was originally designed for the analysis of the
spectra of hot stars with radiatively driven winds, but an earlier version has
already been used by \citet{pauldrach96} (Paper~I) to quantitatively
investigate the effects of line blocking in {\snia}. While they used a
consistent treatment of line blocking, the back-reaction of the line opacities
on the temperature structure (line blanketing) was not taken into account.  In
our present work the current version of the code has been further adapted to
treat the radiative transfer in supernovae in a more sophisticated way.

Section~\ref{sec:model} describes the setup of the non-LTE model and introduces
the numerical scheme used to solve the radiative transfer. In
Section~\ref{sec:photosphere} the physical conditions in the
``pseudo-photosphere'' of {\snia} are discussed with respect to the solution of
the radiative transfer and are compared to the situation in normal stars (i.e.,
stars that have a well defined photosphere).  Section~\ref{sec:bc} describes
the derivation of an improved treatment of the inner boundary for the numerical
solution of the radiative transfer. The results are discussed in
Section~\ref{sec:discussion}, and a comparison of a model spectrum with an
observed {\snia} spectrum is shown.  The conclusions are provided in
Section~\ref{sec:conclusion}.

\section{The radiative transfer model}\label{sec:model}

The code {\em WMbasic} has been successfully used to model extended,
radiation-driven stellar atmospheres, assuming a homogeneous,
stationary, spherically symmetric outflow.  Here we provide a brief
outline of the main concepts; details relevant to the
derivation in Section~\ref{sec:bc} will be discussed as well.  A more
comprehensive description of the numerical methods can be found in
\citet{pauldrach01} and \citet{pauldrach03} and references therein.

\subsection{General setup}

To derive synthetic spectra of supernovae in early phases the analogy to hot
stars with extended atmospheres can be used to apply similar concepts for the
solution.  The radiative transfer model for supernovae requires the following
input:
\begin{itemize}
  \item Hydrodynamic structure --- This defines the relationship between radius
    and density and velocity at a given epoch (i.e., time after explosion). Because
    the expansion is ballistic, a homologous structure of the ejecta is generally
    a safe assumption (to describe properties of the ejecta, the velocity is
    usually used as a radial coordinate because it is independent of the epoch).
    Thus, radius $r$, epoch $t$, and velocity $v$ are related via $r=v t$,
    and a density $\rho_0(r,t_{0})$ given at an epoch $t_{0}$ scales
    with $t$ as $\rho(r,t)=\rho_0((t_{0}/{t})r,t_{0})(t_{0}/t)^{3}$.  The
    relationship between density and radius also sets the optical depth scale,
    which defines the ``photospheric radius.'' The hydrodynamic quantities $r$,
    $v$, and $\rho$ remain fixed during the calculation of the non-LTE model.
  \item Luminosity --- In {\snia} the luminosity results primarily from the
    energy that is deposited by $\gamma$-photons originating from the decay of
    radionuclides synthesized in the explosion.  Therefore, part of the
    luminosity will depend on the radial distribution of the respective elements
    in the ejecta.  At the inner boundary of the computational volume an incoming
    luminosity must be specified to account for the radiative energy deposited
    below that boundary.  This luminosity represents both the instantaneously
    released photons as well as the photons that have been
    generated at earlier epochs and trapped by the large opacities. (See, e.g.,
    \citealt{arnett82,hoeflich96,nugent97,pinto00a}.) At the current stage of our 
    project, the explicit energy deposition within the ejecta has not been taken
    into account; instead, the total luminosity is assumed to originate from below
    the boundary of the computational grid. Also, the luminosity is not
    derived from the mass of {\nifs} predicted by a specific explosion model but
    is considered as an independent input parameter.
  \item Epoch --- For homologously expanding ejecta, the epoch determines the
    absolute radius and density scale. Because the exact time of explosion of an
    observed supernova is generally not known, the epoch represents an important
    fit parameter.
  \item Composition --- The elemental abundances can be either taken from the
    nucleosynthesis calculations (to study the properties of a specific explosion
    model) or adjusted independently to fit an observed spectrum in order to
    determine the abundances of the object.  In general, the composition in
    supernova ejecta is a function of radius.
\end{itemize}

\begin{figure*}[htb]
  \includegraphics[width=12cm]{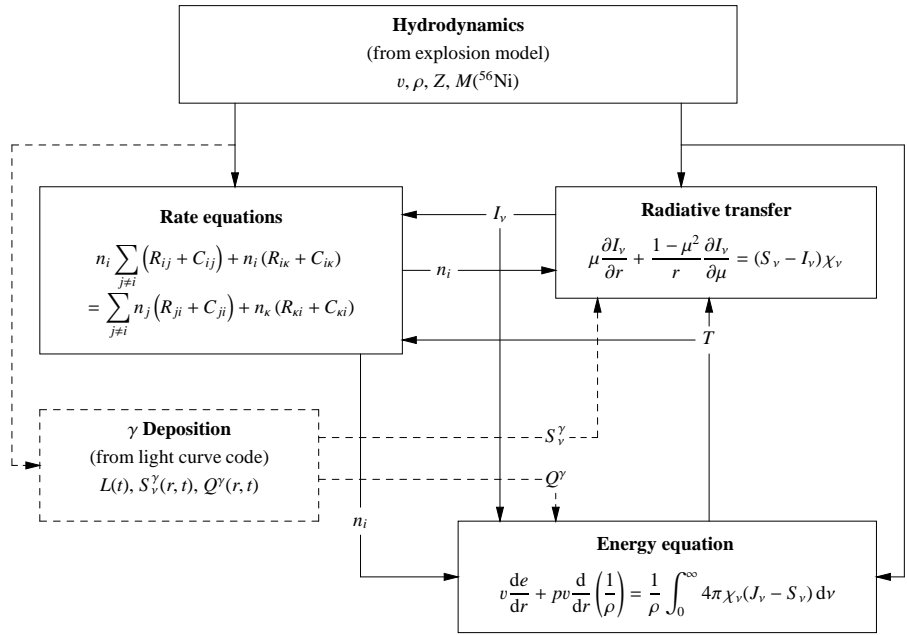}
  \caption{Overview of the physical equations that must be consistently solved
  for the non-LTE model. The input is fixed by an external
  explosion model. In the middle-left box the rate equations which
  determine the occupation numbers $n_{i}$ of the atomic levels are given. The
  middle-right box shows the radiative transfer equation for the radiation
  field. The lower-right box gives the energy equation that fixes the
  temperature structure within the atmosphere. The energy deposition by
  $\gamma$ photons, shown in the lower-left box, is currently taken
  into account only approximately. This means that the models rely on the
  input of the total luminosity $L$ at the lower boundary of the computational grid
  instead of the $^{56}$Ni-mass.}
  \label{fig:nlte-physics}
\end{figure*}

The computation of a self-consistent non-LTE-model requires the simultaneous
solution of a number of physical equations, in particular the radiative
transfer, the rate equations, and the energy equation.
Fig.~\ref{fig:nlte-physics} gives an overview of the physical equations and
the physical quantities by which these equations are interconnected.

The general concept behind our code, {\em WMbasic}, is to first obtain a rough
solution with a fast, approximate method, and then, based on this solution,
obtain a completely consistent solution with an exact, detailed method. The
approximate method should provide a solution that is sufficiently close to the
final solution so that only a few iterations with the much more time-consuming
detailed method are necessary.

Our fast, approximate method is based on a Doppler-broadened {\em sampling
technique} for line opacities and emissivities.  The idea behind this method is
to solve the radiative transfer for a {\em representative} sample of frequency
points in the relevant spectral range.

In the final iterations, a detailed radiative transfer method that does not
suffer from the approximations of the first iteration cycle is used. This
method uses an exact observer's frame solution, equivalent to a comoving frame
solution, which correctly treats the angular variation of line opacities and
emissivities. The line profiles are spatially resolved. Multi-line effects and
back-reactions of the line opacities on the model structures are treated
correctly.

The temperature structure is in practice obtained from balancing energy gains and losses to
the electron gas (heating and cooling rates). This description is equivalent to
the condition of radiative equilibrium (indicated in the energy equation in
Fig.~\ref{fig:nlte-physics}), but is numerically advantageous for physical
conditions where the opacity is dominated by scattering events that do not
couple the radiation field to the thermal pool (see \citealt{pauldrach01}).

\subsection{Solution of the radiative transfer}
\label{sec:rad-trans}

To introduce the nomenclature and equations used later, the solution
applied to solve the radiative transfer in the observer's frame
employing the fast opacity sampling method is recapitulated in this section.

\begin{figure}[htb]
  \includegraphics[width=8.4cm]{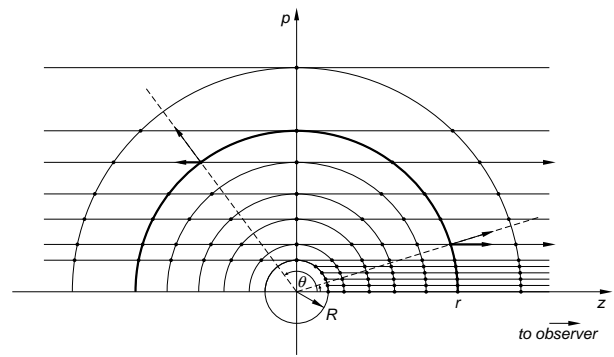}
  \caption{\label{fig:pz-geom}$p$-$z$ coordinate system used to solve the
  transfer equation in spherical symmetry. $R$ denotes the radius of the inner core.}
\end{figure}

To determine the radiation field that enters into the Thomson emissivity, an
iteration alternating between the ray-by-ray solution and the
angular-integrated moments equation is performed.  Both systems are solved with
a Feautrier-type scheme \citep{feautrier64} as discussed, e.g., in
\citet{mihalas78}.  For each frequency point, the iteration is performed twice:
first for a pure continuum model and afterwards for the full problem with
continuum and lines.  The solution is carried out in the usual Cartesian-like
$p$-$z$-coordinate system where each $p$-ray at a given radius shell corresponds to a
$\mu=\cos\vartheta$-direction in spherical coordinates (see
Fig.~\ref{fig:pz-geom}).
The transfer equation for each $p$-ray is
rewritten for the intensities in positive and negative $z$-direction
\begin{equation}
  \label{eq:transfer-z}
  \diff{I^{\pm}_{\nu}}{\tau_{\nu}} = \pm(S_{\nu}-I^{\pm}_{\nu}) \quad \mr{with}\quad
  \dd\tau_{\nu} = -\chi_{\nu} \dd z.
\end{equation}
These equations have to be evaluated at each frequency point of the computational
grid.  In the remainder of this section the index $\nu$ is omitted for brevity.
By introducing the new variables $u = \half\left(I^{+}+I^{-}\right)$ and
$v=\half\left( I^{+}-I^{-} \right)$ these two first order differential
equations with a single boundary condition (initial value problems) can be
converted into a second order differential equation with a boundary condition
for each side:
\begin{equation}\label{eq:feaut3}
  \diff{u}{\tau}= v, \quad \diff{v}{\tau}= u - S \quad
 \Ra \quad \sdiff{u}{\tau} = u - S
\end{equation}
where $S$ is the source function.

To close the system, suitable boundary conditions must be specified.
At the outer boundary the condition $I^{-}\equiv 0$ (no radiation
incident from outside) leads to
\begin{equation}
  u = v \quad \Ra \quad \left.\diff{u}{\tau}\right|_{r=r_{\rm max}}=u.
\end{equation}
At the inner boundary one has to distinguish between those $p$-rays that
intersect the photospheric core ($p<R$) and those that do not ($p>R$). (See
Fig.~\ref{fig:pz-geom}.)  For core rays the incident intensity has to be
explicitly specified, $I^{+}=I_{\rm core}$, while for non-core rays a
reflecting boundary $I^{+}=I^{-}$ is used.  Noting that $v=I^{+}-u$, one gets
\begin{equation}\label{eq:feaut-core}
  \left.\diff{u}{\tau}\right|_{\tau_{\rm max}}=I_{\rm core} - u \qquad(p<R)
\end{equation}
for core rays and
\begin{equation}
  \left.\diff{u}{\tau}\right|_{\tau_{\rm max}}=0\qquad(p>R)
\end{equation}
for non-core rays.  Using a suitable discretization scheme (e.g., the standard
scheme as modified by \citet{pauldrach01}; see also Hoffmann et~al.\ (2006), in
prep.), the numerical solution of this system can be carried out very efficiently.

A very similar method can be used to obtain a solution for the moments equation
in spherical symmetry
\begin{eqnarray}\label{eq:momrt3}
  \diff{\hti}{\tti}=\frac{1}{q}\left(\jti-\sti\right), \quad
  \diff{(qf\jti)}{\tti}=\hti\label{eq:momrt2} \nonumber\\
  \Ra\quad\sdiff{(qf\jti)}{\tti} = \frac{1}{q}\left(\jti-\sti\right),
\end{eqnarray}
where $J=\int_{-1}^1 I(\mu)\,\dd\mu$, $H=\int_{-1}^1 I(\mu)\,\mu\,\dd\mu$,
and all symbols with a tilde denote the respective quantity times $r^{2}$.
The Eddington factor $f_{\nu}$ is defined as
\begin{equation}
  f=\frac{J}{K}=\frac{\int_{-1}^1 I(\mu)\,\dd\mu}{\int_{-1}^1 I(\mu)\,\mu^2\,\dd\mu}
  \label{eq:fnu}
\end{equation}
(at each frequency point $\nu$);
$q$ denotes the sphericality factor, defined by
\begin{eqnarray}
  &&\diff{(r^{2}q)}{r} :=
  r^{2}q\frac{3f-1}{rf}\\
  &&\Ra\quad
  qr^{2}=\exp\left({\int_{1}^{r}\frac{3f(r')-1}{r'f(r')}\,\dd r'}\right).
  \label{eq:sphq}
\end{eqnarray}

One advantage of the moments equation, as compared to the ray-by-ray
solution, is that one can implicitly solve for the contribution of Thomson
scattering to the source function $S$.  Separating the emissivity due to true
processes ($\eta^{\rm true}$) and that due to Thomson-scattering ($\eta^{\rm Th} = \chi^{\rm Th}J$),
one can write
\begin{equation}
  S = \frac{\eta^{\rm true}}{\chi^{\rm true}+\chi^{\rm Th}} +
  \frac{\chi^{\rm Th} J}{\chi^{\rm true}+\chi^{\rm Th}} =
  (1-\beta)S^{\rm true} + \beta J
\end{equation}
with the definitions
\begin{equation}
  \label{eq:stru}
  S^{\rm true}:=\frac{\eta^{\rm true}}{\chi^{\rm true}}\qquad\mr{and}\qquad
  \beta:=\frac{\chi^{\rm Th}}{\chi^{\rm true}+\chi^{\rm Th}}.
\end{equation}
Using this in \req{eq:momrt3} gives
\begin{equation}
  \label{eq:feautm2}
    \sdiff{(qf\jti)}{\tti} =
    \jti(1-\beta)-\sti^{\rm true}(1-\beta).
\end{equation}
At the boundaries, the system is closed by employing factors
\begin{equation}
  \label{eq:eddh0}
  h  :=  \frac{\int_{0}^{1}u(\mu)\,\mu\,\dd\mu}{\int_{0}^{1}u(\mu)\,\dd\mu}
\end{equation}
similar to
the second Eddington factor\footnote{The second Eddington factor is
actually defined as the ratio $\left.(H/J)\right|_{r=r_{\rm max}}$ at the outer
boundary of the atmosphere.},
with $u(\mu)$ coming from the solution of the ray-by-ray solution.  At
the outer boundary, because $u(r_{\rm max})\equiv v(r_{\rm max})$, this is
\begin{equation}\label{eq:eddh1}
  h(r=r_{\rm max}) = \left.\frac{H}{J}\right|_{r=r_{\rm max}}.
\end{equation}
Thus, the outer boundary equation is
\begin{equation}
  \left.\diff{(fq\jti)}{\tti}\right|_{r=r_{\rm max}} = h(r=r_{\rm max})\jti(r=r_{\rm max}).
\end{equation}
The inner boundary ($r=R$) is treated similarly; however, because $\int
u\,\mu\,\dd\mu \ne H$, $I_{\rm core}$ from the ray-by-ray solution has
to be employed here as well, noting that
\begin{equation}
  H(\tau_{\rm max}) = \int_{0}^{1}\!\! v\,\mu\,\dd\mu
  = \int_{0}^{1}\!\!I_{\rm core}\,\mu\,\dd\mu -
  \int_{0}^{1}\!\!u\,\mu\,\dd\mu.
\end{equation}
This results in
\begin{equation}\label{eq:feautm-bci}
  \left.\diff{(fq\jti)}{\tti}\right|_{\tau_{\rm max}} =
  R^{2}\int_{0}^{1}\!\!I_{\rm core}\,\mu\,\dd\mu - h(\tau_{\rm max})\jti.
\end{equation}

In the final, detailed solution, the radiative transfer is solved through the
total hemisphere using a formal integral on an adaptive micro-grid. This also
requires the specification of boundary conditions. For consistency, the
$I^{+}_{\rm core}$ values used for the sampling method must be used for the
core rays in the iterations using this method. Therefore, only
the boundary condition for the sampling iteration will be discussed here.

\section{Physical conditions at the photosphere of~a~{\snia}}
\label{sec:photosphere}

In normal stellar atmospheres the exponential increase of the density at the
bottom of the atmosphere provides a clear definition of a photospheric radius
because large optical depths are reached at all wavelengths within a very short
spatial distance.  In contrast, {\snia} do not have a clearly defined
photosphere:  because the material is unbound, no exponential density structure
comparable to a stellar atmosphere can develop.  As a result, the optical depth
scale depends strongly on the wavelength. (This makes the concept of a mean
optical depth like the Rosseland optical depth much less useful, if not
entirely meaningless, in {\snia}.) In addition, the absolute densities are much
lower than in stellar atmospheres and the composition in {\snia} is dominated
by intermediate-mass and iron-group elements. When compared to stellar
atmospheres this behavior leads to very low number densities of ions and
electrons, resulting in a significantly weaker free-free continuum. The absence
of hydrogen and helium in the ejecta further reduces the contribution of the
bound-free continuum in the optical and infrared part of the spectrum.  Going
from red to blue wavelengths the first strong continuum edge is the \ion{O}{i}
ionization edge at $911\,${\AA}.

\begin{figure}[htb]
  \centering \includegraphics[width=8.4cm]{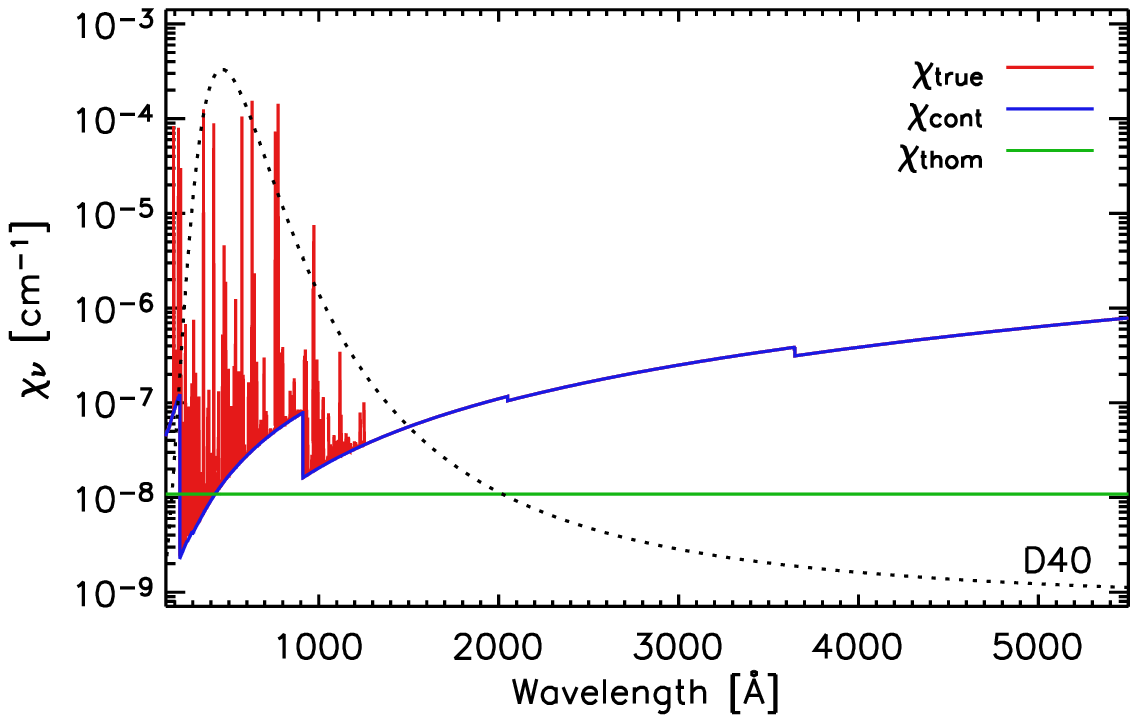}
  \includegraphics[width=8.4cm]{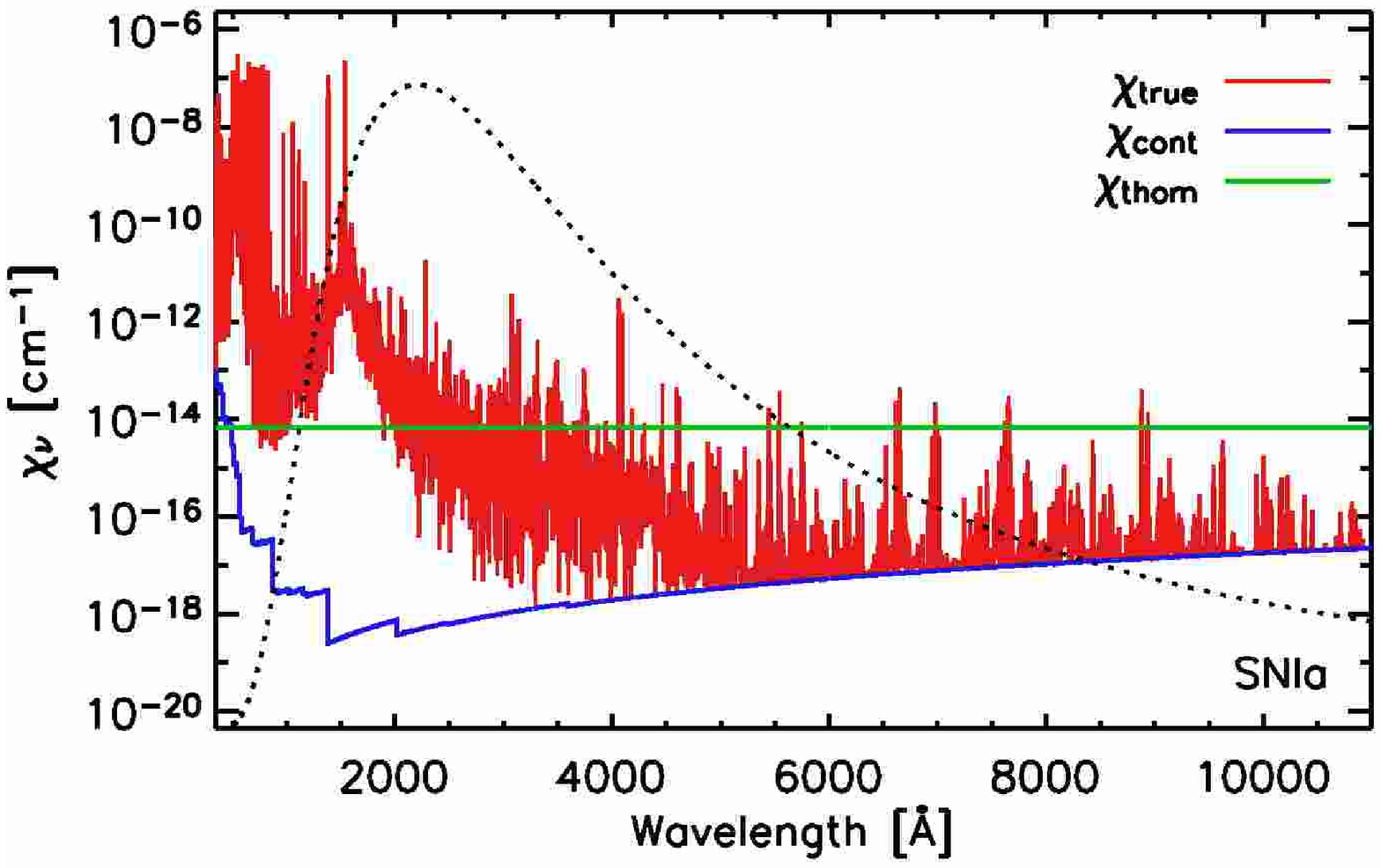}
  \caption{\label{fig:opaccomp}
  Comparison of the different contributions to the total opacity near the
  photosphere in a stellar atmosphere (upper panel; model D40 in
  \citealt{pauldrach01}) and a {\snia} model (lower panel); note the
  logarithmic scale. $\chi_{\rm true}$ represents the total true opacity
  (continuum and lines), $\chi_{\rm cont}$ the true continuum alone and
  $\chi_{\rm thom}$ the Thomson scattering opacity. The dotted lines show the
  shape of a blackbody spectrum at the respective temperature on an arbitrary
  scale to indicate the position of the flux maximum.}
\end{figure}

Fig.~\ref{fig:opaccomp} shows a comparison of the contributions to the total
opacity in the photospheric region of a hot star (Model D40 from
\citet{pauldrach01}) and a {\snia}.  The dotted line shows a blackbody spectrum
corresponding to a typical temperature ($40\,000\,$K for D40, $12\,000\,$K for
the supernova) to indicate the approximate position of the maximum flux in the
spectrum. It can be seen that in the supernova, the bound-free continuum
opacity is irrelevant compared to the line opacity in the major part of the
spectrum.  The plot also illustrates the formation of the ``pseudo-continuum''
by the overlap of thousands of lines.

\begin{figure}[htb]
  \centering
  \includegraphics[width=8.4cm]{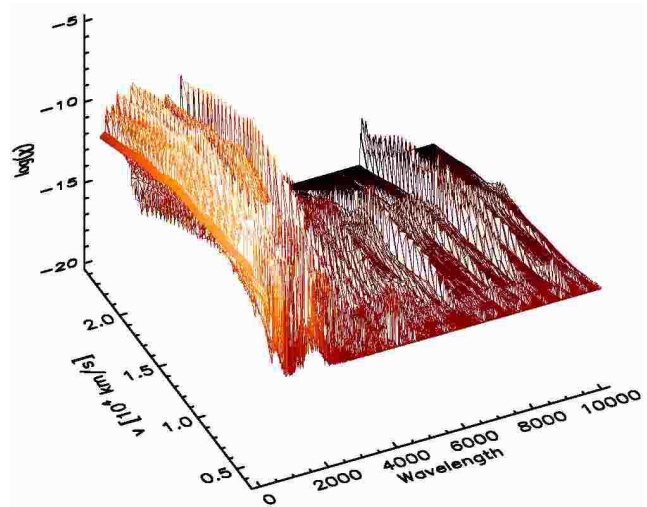}
  \caption{Logarithm of the total opacity in a {\snia} model (sampling
  iteration) versus velocity and wavelength. Note that the line
  opacity {\em decreases} toward the inside (front) because higher
  ionization stages with less lines dominate.}
\label{fig:opa3dsnia}
\end{figure}

The most significant qualitative difference between a supernova and a star,
however, is that at wavelengths redward of about $5000\,${\AA}, the electron
scattering opacity becomes the dominating source of opacity, even in deep
layers of the ejecta. Note that this situation cannot be changed significantly
by computing down to smaller radii because the mild increase of density does
not permit the formation of a significant free-free opacity. On the contrary,
one encounters a lower contribution of true opacities because higher ionization
stages tend to have less lines, and thus the line opacity decreases inwards.
(In addition, the validity of the stationarity assumption needs to be
considered because the photon trapping time at deep layers becomes comparable
to the escape time.) Fig.~\ref{fig:opa3dsnia} shows the logarithm of the
total opacity as a function of velocity and wavelength for a {\snia} model
(epoch: $25\,$days after explosion) where the effect of decreasing line opacity
can be seen clearly.

Compared to a blackbody spectrum at the respective temperature, the radiation
field in the innermost regions is  more likely to have a bluer characteristic
because it is the result of radiation from the deposition of $\gamma$ rays that
is not entirely thermalized.  Furthermore, no emission from down-scattering of
$\gamma$-photons can be generated further out in the ejecta in  wavelength
regions that do not have significant continuum or line opacity. This effect
results in the characteristic shape of {\snia} spectra in red and infrared
wavelengths where the slope of the pseudo-continuum is generally steeper than
the slope of a corresponding blackbody spectrum. In synthetic spectra computed
assuming thermalization this effect generally results in an offset of the model
spectrum in the red and infrared wavelengths (see e.g., \citealt{pauldrach96}
and \citealt{nugent97} and the spectral fits in \citealt{stehle05}).

In summary, the ejecta of early {\snia} form an intermediate object between an
extended stellar atmosphere and a planetary nebula. For both extreme cases the
choice for the boundary conditions is clear: for the star, the LTE diffusion
approximation  \req{eq:lte-da} is a suitable choice.  For a gaseous nebula, the
incident radiation field from the illuminating star naturally defines the
radiation field at the inner boundary (because there is essentially no
back-reaction of the nebula on the stellar atmosphere). In {\snia} neither of
these choices can be strictly applied.

In the next section we will discuss an extension to the diffusion approximation
that eliminates most of the restrictive requirements of LTE and therefore
allows a more consistent description of the inner boundary for supernova
conditions.

\section{An improved inner boundary for the radiation transfer at non-LTE
  conditions}
\label{sec:bc}

\subsection{The standard case for stellar atmospheres:
  the~LTE diffusion approximation}
\label{sec:da-std}

The numerical solution of the radiative transfer equation requires a set of
boundary conditions.  For the objects discussed here, this requires an
assumption about the incoming radiation $I_{\nu}^{+}$ at the core.
$I_{\nu}^{+}$ should be chosen so that it describes the radiation field  as
accurately as possible under the physical conditions present. Ideally, the
expression for the boundary equation is an analytic extrapolation of the
radiation field at the innermost points.

For the calculations done here, the incident radiation from outside the ejecta
is assumed to be zero.  Therefore, the boundary condition for the outer
boundary is $I^{-}(\nu,\mu)|_{r=r_{\rm max}}=0$. The common choice for the
$I_{\nu}^{+}$ at the inner boundary in stellar atmospheres is derived from  the
LTE-diffusion approximation (see, e.g., \citealt{mihalas78})
\begin{equation}
  \label{eq:lte-da}
  I^{+}(\nu,\mu) \approx B_{\nu}(T) + 3\mu \diff{B_{\nu}(T)}{\tau_{\nu}}.
\end{equation}
This result is more generally obtained by applying a harmonic expansion for
$I^{+}$ that leads to the expression (see, e.g., \citealt{pomraning73})
\begin{equation}\label{eq:diffapp0a}
  I^{+}(\nu,\mu) \approx I^{0}(\nu) +
  3\mu I^{1}(\nu)\equiv I^{0}_{\nu} + 3\mu H^{0}_{\nu}
\end{equation}
where the zeroth term $I^{0}_{\nu}$ is isotropic and the first term has an
angular dependence proportional to $\mu$. The term $I^{1}$ has the properties
of a flux (it is, however, in general not identical to the real flux
$H_{\nu}$ that results from a solution of the transfer equation with
this boundary condition). We therefore associate the symbol
$H^{0}_{\nu}\equiv I^{1}(\nu)$ with the
analytical expression for the anisotropic term of the boundary equation.
In the following we will discuss expressions of different refinement for the
terms $I^{0}_{\nu}$  and $H^{0}_{\nu}$.

For the classical derivation of the LTE diffusion approximation, commonly used
in stellar atmospheres, a Taylor expansion of $S_{\nu}$ in the limit of large
$\tau$
\begin{equation}\label{eq:diffapp0}
  S_{\nu}(\tau_{\nu}') \approx \bnu = \sum_{n=0}^{\infty}
  \frac{\partial^{(n)}\bnu(\tau_{\nu})}{\partial\tau_{\nu}^{(n)}}
  \frac{(\tau_{\nu}'-\tau_{\nu}^{n})}{n!}
\end{equation}
is used. Through the formal solution of the transfer equation this leads to
the terms
\begin{equation}\label{eq:diffapp1}
  I^{0}_{\nu} \approx I^{0}_{\rm classic}(\nu)\equiv B_{\nu}(T)
\end{equation}
and
\begin{equation}\label{eq:diffapp2}
  H^{0}_{\nu} \approx H^{0}_{\rm classic}(\nu) \equiv
  \frac{1}{3}\diff{B_{\nu}(T)}{\tau}  =
  -\frac{1}{3}\frac{1}{\chi_{\nu}}\pdiff{\bnu}{T}\diff{T}{r}.
\end{equation}
In the standard implementation, the (lower) boundary equations for the
ray-by-ray solution (core rays), \req{eq:feaut-core} and the moments equation
\req{eq:feautm-bci}, are therefore given by
\begin{eqnarray}
  \left.\diff{u_{\nu}}{\tau_{\nu}}\right|_{\tau_{\nu}^{\rm max}} &=& I_{\nu}^{0} + 3\mu H_{\nu}^{0}
  - u_{\nu}(\tau_{\nu}^{,\rm max})\\
  \left.\diff{(f_{\nu}q_{\nu}\jti_{\nu})}{\tti_{\nu}}\right|_{\tau_{\nu}^{\rm max}} &=&
  \left(\half I_{\nu}^{0} + H_{\nu}^{0}\right)R^{2} - h_{\nu}(\tau_{\nu}^{\rm
    max})\jti_{\nu}(\tau_{\nu}^{\rm max}).
\end{eqnarray}

The expansion \req{eq:diffapp0} and, therefore, \req{eq:diffapp1} is applicable
if the radiation field is thermalized (i.e., the mean free paths of photons are
much shorter than any significant hydrodynamic length scale).  In {\snia},
however, this condition is not fulfilled over the full spectrum (as discussed
in Section~\ref{sec:photosphere}). Hence, the use of \req{eq:diffapp1} leads to
incorrect spectral properties of the radiation field at the inner boundary.
The inconsistency between the radiation field and the physical state of the
matter caused by enforcing a thermal radiation field at the core boundary leads
to spurious results in the rate equations and, in particular, in the heating
and cooling rates for the temperature determination. The unstable behavior
results from a feedback effect between the radiation field and the temperature.
The radiation field at the inner points is used to determine the temperature,
which (through the boundary equation) sets the incoming intensity and,
therefore, partially sets the radiation field at those points.

Part of this work focuses on deriving an analytical expression for the
radiation field at the inner boundary that reflects the physical conditions in
{\snia} more accurately and reproduces the slope of the pseudo-continuum in
the red and infrared wavelengths better.

\subsubsection{Flux constraint}

To constrain the total flux at the inner boundary, the frequency integrated
input flux $H^0 =\int_{0}^{\infty}\!\!H^{0}_{\nu}\,\dd\nu$ is compared to the
total integrated input flux ${\cal H}_{0} =L/(16\pi^{2}R^{2})$. This results in
a (frequency-independent) scaling factor
\begin{equation}
  \label{eq:fluxcorr}
  FC = \frac{{\cal H}_{0}}{\int_{0}^{\infty}\!\! H^{0}_{\nu}\,\dd\nu}
\end{equation}
that is applied to the flux term of the inner boundary condition,
so that (cf.\ Eq.~\ref{eq:diffapp2})
\begin{equation}
  H^{0}_{\rm classic{'}}(\nu) =
  -\frac{1}{3}\frac{1}{\chi_{\nu}}\pdiff{\bnu}{T}\diff{T}{r} FC_{\rm classic}
  \label{eq:H-FC}
\end{equation}
  with
\begin{equation}
  FC_{\rm classic} =
  \frac{\frac{L}{16\pi^{2}R^{2}}}{-\diff{T}{r}\int_{0}^{\infty}\!\!
    \frac{1}{\chi_{\nu}}\diff{B_{\nu}(T)}{T}\,\dd\nu}
\end{equation}
(\citealt{mihalas78}, p.~252). The purpose of applying this flux correction
factor is to induce the model to converge to a solution with the correct
total flux. The factor $FC$ not only accounts for small numerical differences
(in particular in the derivation of the temperature gradient) but also for
inconsistencies between the assumptions implied in the construction of the
boundary flux and the physical conditions that are actually encountered at
the boundary layer of the model. For the case of the classical diffusion
approximation, deviations from $FC=1$ in the converged model indicate
departures from ideal LTE diffusion conditions.

With respect to the moments equation for the flux at the inner boundary the
derived flux is actually
\begin{equation}\label{eq:hnuib}
  H_{\nu}(R) = \!\!\int_{0}^{1}\!\!\left(I^{0}_{\nu}+3\mu
    H_{\nu}^{0}\right)\mu\,\dd\mu -
  h_{\nu}J_{\nu} = H_{\nu}^{0} + \left(\half I^{0}_{\nu}-h_{\nu}J_{\nu}\right).
\end{equation}
Thus, constraining the flux by adjusting $H^{0}_{\nu}$ only with respect to
$H^{0}$ implicitly assumes that the second term in \req{eq:hnuib} vanishes,
which requires
\begin{equation}
  h_{\nu}(R)J_{\nu}(R) = \half I^{0}_{\nu}.
\end{equation}
\begin{figure}[htb]
  \centering
    \includegraphics[width=8.4cm]{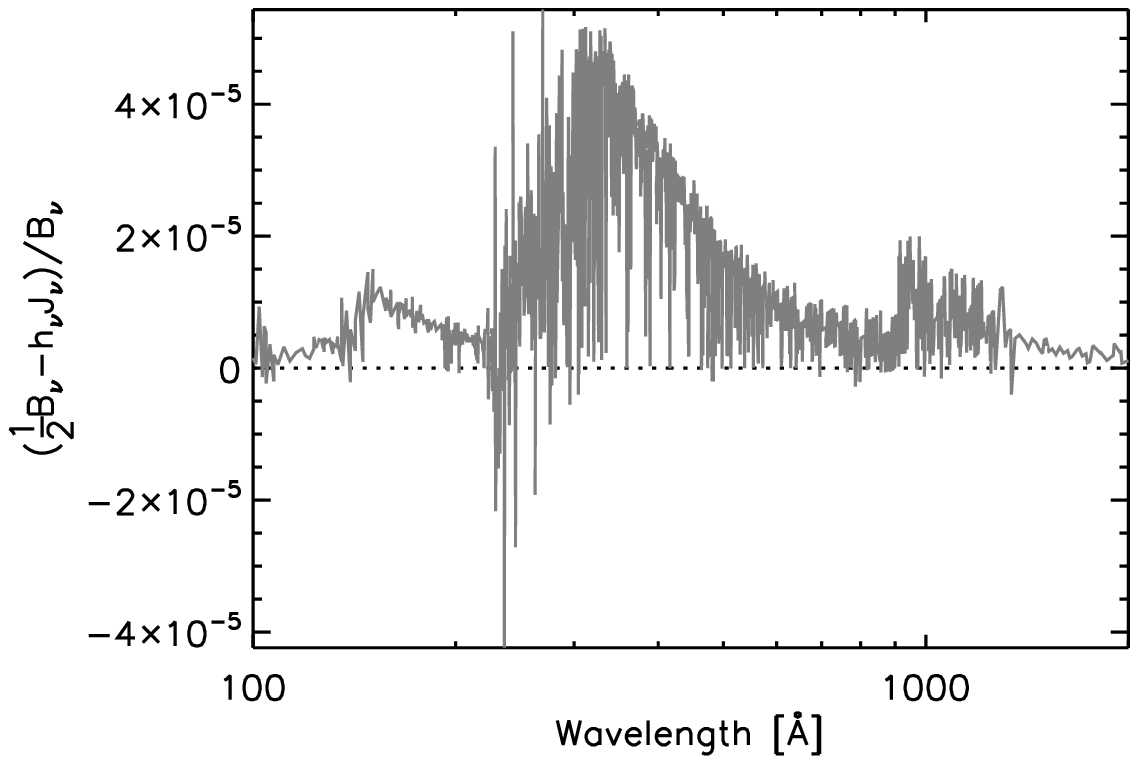}\\
    \includegraphics[width=8.4cm]{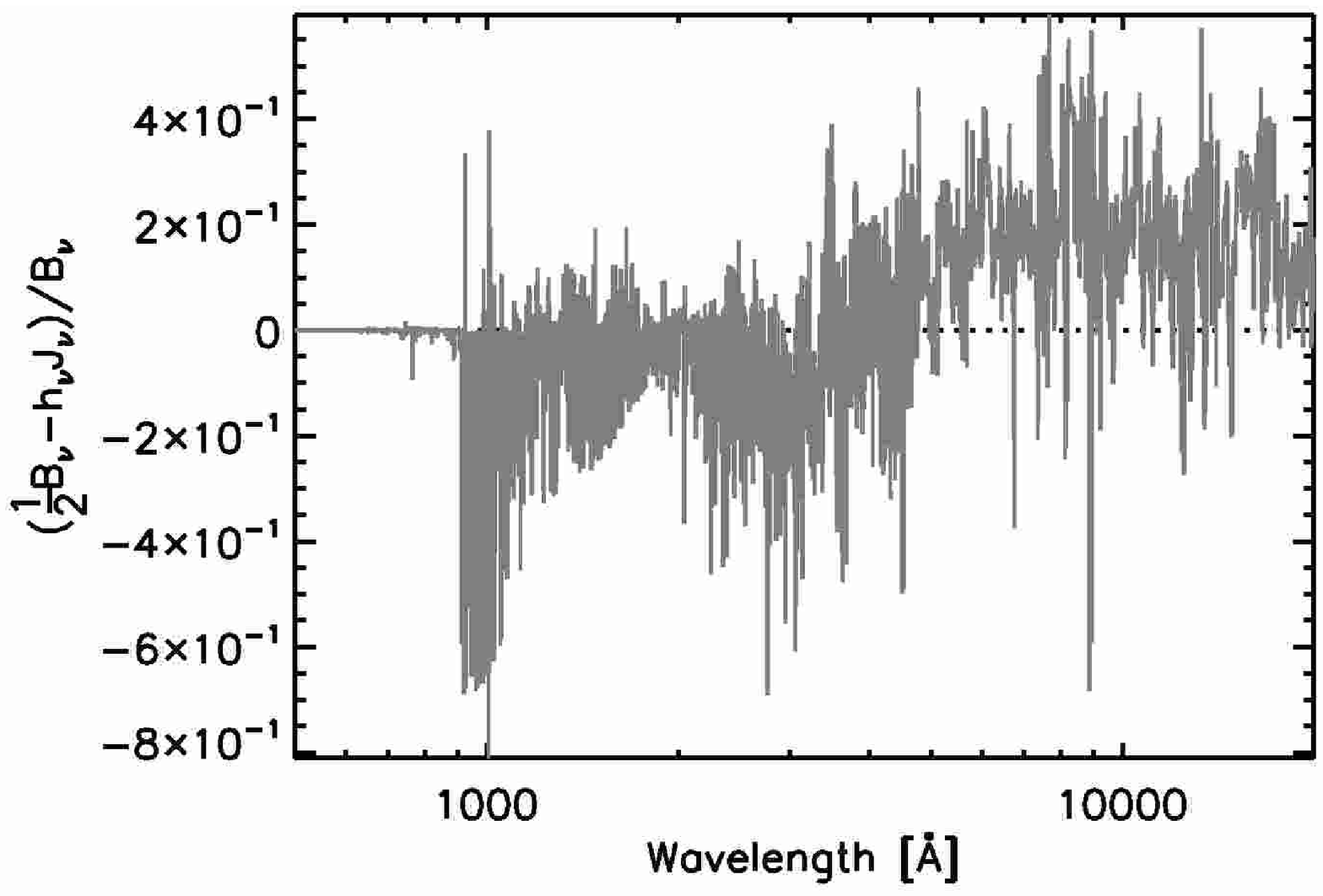}
    \caption{The ratio  $(\half\bnu-h_{\nu}J_{\nu})/\bnu$ at the inner boundary
      (see text). {\em Upper panel:} O-star model D40 \citep{pauldrach01}.
      {\em Lower panel:} {\snia} model. Note the different $y$-scales in
      these plots.}
  \label{fig:hib}
\end{figure}
As can be seen in Fig.~\ref{fig:hib}, even for the D40 star model (cf.\
\citealt{pauldrach01}) this condition is not exactly fulfilled.  In this case,
however, the deviations are certainly negligible, whereas in {\snia} the effect
is much stronger (bottom panel in Fig.~\ref{fig:hib}). This leads to a
significant discrepancy between $FC$ and the actually derived flux. The
modifications discussed in the next section require that this effect is taken
into account.  Therefore, the correct flux correction factor has to be used:
\begin{equation}
  \label{eq:fluxcorrnew}
  FC = \frac{\frac{L}{16\pi^{2}R^{2}} -  \int_{0}^{\infty}\left(\half
      I^{0}_{\nu}-h_{\nu}J_{\nu}\right)\,\dd\nu }{\int_{0}^{\infty}\!\!
    H^{0}_{\nu}\,\dd\nu}.
\end{equation}
Without the additional term that accounts for small deviations from LTE the
integrated flux at the inner boundary deviates from the desired input flux even
if $FC=1$.
With the formulation in \req{eq:fluxcorrnew}, it is possible to achieve $FC=1$ {\em and} the
correct flux at $r=R$.

As already noted above, in the formulation of Eq.~\ref{eq:H-FC} the factor $FC$
effectively represents a correction to the temperature gradient at the inner
boundary:
\begin{equation}
  I^{+}(\nu,\mu) = B_{\nu}(T) +
  \mu\frac{1}{\chi_{\nu}}\pdiff{\bnu}{T}\left(FC\diff{T}{r}\right).
  \label{eq:dafc1}
\end{equation}
We have found that it is numerically more stable to determine the temperature
$T_1$ of the innermost grid point directly from requiring that the correct flux
be reached, instead of applying $FC$ as above and waiting for the
temperature-correction mechanism of the code (based on heating and cooling
rates) to adjust the temperature accordingly.  One reason for an unstable
behavior of the latter approach is that in \req{eq:dafc1} the first term is
also temperature dependent, and, in case of the physical conditions in a
supernova, is of the same order as the second term. Therefore, a correction of
the second term alone can lead to an inconsistency between the shape of the
radiation field that results and the actual temperature.  (An additional
problem under near-LTE conditions (e.g., at the inner boundary in ``normal''
stellar atmospheres) is that determining the temperature from the balance of
the kinetic heating and cooling rates becomes inaccurate, because in thermal
equilibrium the heating and cooling rates balance at all temperatures.  In
stellar atmosphere models we avoid the problem by computing the temperature in
the optically thick inner region either via the condition of radiative balance,
or as a parametrized function of the optical depth, with the parameters
adjusted to conserve the flux (cf.~\citealt{pauldrach01}).)  From
\req{eq:fluxcorrnew} and the condition $FC=1$ the temperature of the innermost
point is derived by requiring
\begin{equation}
  \int_{0}^{\infty}\!\!H^{0}_{\nu}\,\dd\nu-\left(\frac{L}{16\pi^{2}R^{2}} - \!\!
     \int_{0}^{\infty}\!\left(\half I^{0}_{\nu}-h_{\nu}J_{\nu}\right)\,\dd\nu\right)
     \stackrel{!}{=}0
  \label{eq:t1}
\end{equation}
where, for the case of the standard diffusion approximation, $H^{0}_{\nu}$
corresponds to the flux term $H^{0}_{\rm classic}(\nu)$ of the boundary
condition (\req{eq:diffapp2}).  The conditional equation \req{eq:t1}
implicitly defines $T_1 = T_2 + (\dd T/\dd r) \Delta r_{1,2}$ because
$I^{0}_{\nu}$ and $H^{0}_{\nu}$ are functions of $T$ and $\dd T/\dd r$
(Eqs.~\ref{eq:diffapp1} and \ref{eq:diffapp2}).  The temperature $T_2$ of the
second grid point (and all other grid points) is derived conventionally via
heating and cooling rates.  In practice,  \req{eq:t1} is solved numerically
in each iteration, assuming that the temperature-dependence of the second
integral is negligible.

Note that the expression \req{eq:hnuib} implicitly contains the assumption of
thermalization. Only then is the degree of isotropy given that is necessary to
make the expansion \req{eq:diffapp0a} meaningful, neglecting quadratic and
higher order terms.

\subsection{$I^{+}$ for a non-thermal radiation field at the inner~boundary}
\label{sec:iplus}

In this section we will consider modifications to the inner boundary that allow
deviations of the radiation field from thermal equilibrium conditions, which
reflects the physical situation in {\snia} better.  A consistent treatment of
the boundary becomes increasingly important for models of later epochs, as long
as the luminosity emitted at the boundary is still significant compared to the
flux originating from the $\gamma$-ray energy deposition above that boundary.

All modifications must be carried out in such a way that, in the limit of
LTE-conditions at the inner points, the standard diffusive boundary condition
\req{eq:lte-da} is retained.  While it will not be possible to determine a
boundary condition entirely free of analytical approximations (the model would
not be sufficiently constrained) some of the assumptions entering into
\req{eq:diffapp1} can be relaxed without affecting the stability of the solution.

Starting from \req{eq:diffapp0a} one can give up the assumption of strictly
thermal conditions by allowing deviations of the terms $I^{0}(\nu)$ and
$I^{1}(\nu)$ from the Planck function. Thus, instead of \req{eq:lte-da} we set
more generally
\begin{equation}
  I^{+}_{\nu}=J^{0}_{\nu} + 3\mu H^{0}_{\nu}
\end{equation}
with an intensity term $J^{0}_{\nu}$ and a flux term $H^{0}_{\nu}$ to be
determined.  We will consider those two terms separately because we will choose
a fundamentally different treatment.

\subsubsection{The isotropic term $J_{\nu}^{0}$}

We have found that the isotropic term $J^{0}_{\nu}$ can be determined
numerically by a simple iteration procedure. This does not involve very much
additional computational effort compared to the standard solution for the
boundary condition because the iteration is carried out to determine the
intensity term for the Thomson scattering source function anyway.

In the solution of the moments equation, one can solve for this term implicitly
by writing the boundary equation as
\begin{eqnarray}
  \diff{(f_{\nu}q_{\nu}\jti_{\nu})}{\tti_{\nu}}\!\!\!&=&
  \half \jti_{\nu} + H_{\nu}^{0}R^{2} - h_{\nu}\jti_{\nu}\\
\Ra\qquad\diff{(f_{\nu}q_{\nu}\jti_{\nu})}{\tti_{\nu}} \!\!\!&-&
\left(\half-h_{\nu}\right)\jti_{\nu}= H_{\nu}^{0}R^{2} \qquad (\tau_{\nu}=\tau_{\nu}^{\rm max}).
\end{eqnarray}
An equivalent expression for the ray-by-ray solution is not as straightforward
and may lead to slightly inconsistent results for $J_{\nu}$ in the iteration
because there is no constraint that requires the angular integral of the
converged analogous quantities $u_\nu$ for each ray to equal $J_\nu$. For the
boundary equation in the ray-by-ray solution we therefore also use the $J_{\nu}$
that results from the moments equation.

Compared to the standard diffusion approximation this effectively means that
the inner boundary is less strongly constrained, which may lead to numerical
instability. To ensure that the boundary condition is still well behaved and to
understand its general behavior we have studied this modification on a simple toy
model before applying it to the radiative transfer code.

\begin{figure}[htb]
    \includegraphics[width=8cm]{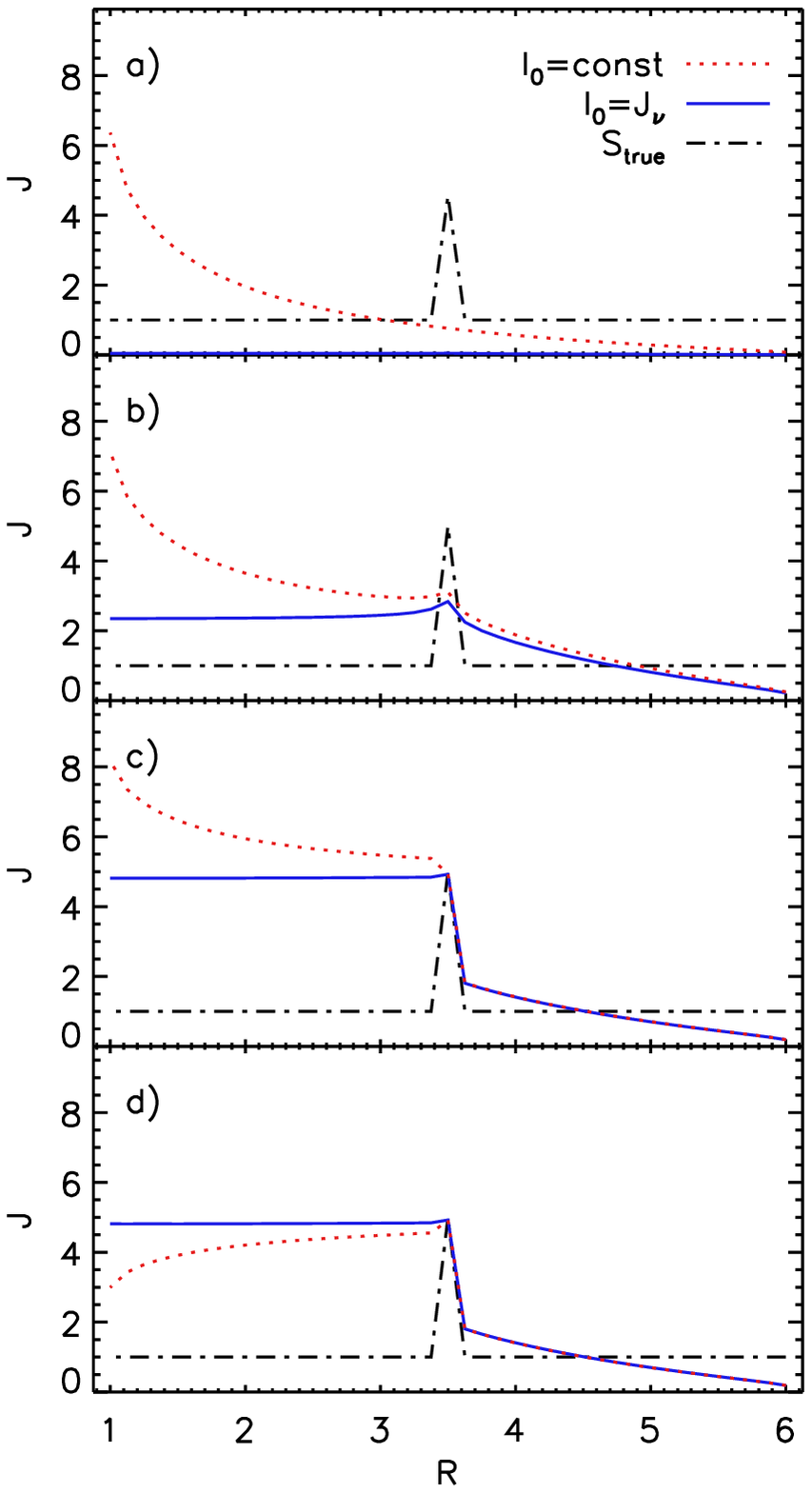}
    \caption{\label{fig:bc-1}Simple toy model to illustrate the new treatment
    of the zeroth term of $I^{+}$ at the inner boundary for the situation of an
    optically thin true continuum. The dotted line represents the run of $J$ if
    the traditional setting $I^{0}=\const$ is assumed, the solid line shows the
    new treatment (see text). The true source function is indicated by the
    dash-dotted line.  Panel {\bf a} shows the case of an optically thin model
    (line and continuum) where $\tau_{\rm line}=0.001$.  Panel {\bf b} shows an
    intermediate case ($\tau_{\rm line}=0.2$).  Panel {\bf c} and {\bf d} show
    the case of an optically thick line ($\tau_{\rm line}=5$) where the
    modification only influences the inner part because the emergent radiation
    is entirely separated from the inner region by the line.  Panel {\bf d} shows a
    situation where the method would overestimate $J$ in the inner region:
    the case where in the conventional treatment $I^{0}$ at the inner boundary
    is smaller than $S_{\rm line}$. In this
    case the iteratively determined $I^{0}$ is not appropriate,
    and instead an $I^{0}$ based on that of a neighboring frequency point
    is used.}
\end{figure}

The results of this toy model are shown in Fig.~\ref{fig:bc-1}.  The basic
parameters are the same for all cases considered. The Thom\-son-opacity is
$\chi_{\rm Th}=1$ and the background opacity (continuum) is $\chi_{\rm c}=10^{-3}$
with a corresponding source function of $S_{\rm c}=1$.
We also consider a line with a source function of $S_{\rm l}=5$
at a certain radius point. (The idea here is that the line is
Doppler-shifted by the velocity field and appears at the
frequency point being considered at that particular radius point.)
The zeroth term of the traditional boundary
condition is set to $I^{0}=10$ for the first three cases and to $I^{0}=2$ for the
last.  $H^{0}$ is set equal to $0$ in all cases. The radiation field is
obtained by an iteration of a ray-by-ray solution with a solution of the
moments equation in spherical symmetry. A Feautrier scheme similar to the one
in the radiative transfer code is used.  Generally, in the main code,
convergence is obtained within less than $15$ iterations -- depending on the
physical conditions and on the relevance of Thomson-scattering in particular.
(For comparison, the iteration for Thomson-scattering alone with a fixed
boundary usually converges within 5 iterations.)  All plots show the comparison
of $J$ as a function of radius obtained from the solution of the moments
equation.  The result from the traditional choice of using a pre-specified $I^{0}=\const$ (in
practice $B_{\nu}$) is shown as a dotted line in red. The result of allowing
$I^{0}$ to consistently converge to $I^{0}=J$
is shown as a solid blue line. The black dash-dotted line represents the true
source function (line and background).

Panel {\bf a} in Fig.~\ref{fig:bc-1} shows the situation for an optically thin
model ($\tau_{\rm line}=0.001$). Panel {\bf b} shows an intermediate case with
$\tau_{\rm line}=0.2$. The last two panels, {\bf c} and {\bf d}, show the case
of an optically thick line ($\tau_{\rm line}=5$). Here the modification only
influences the radiation field in the inner region as the emergent radiation is
entirely separated from the inner region by the optically thick line.
Compared to the conventional treatment with a small $I^{0}$,
in the new treatment $J$ is significantly larger inner region. The
physical conditions cannot cause this increase in intensity because of the low
true opacity.  Thus, in this situation a shortcoming of this method
becomes apparent: one would expect $J_{\nu}$ to drop to $\half S_{\rm l}$
toward the inner boundary because the absence of emission toward the inner
region means that $I^{+}\equiv 0$. The model, however, effectively sets
$I^{+}=I^{-}$.  Unfortunately, this situation occurs quite frequently in
{\snia}:  at each frequency point where the opacity at the boundary is
low but increases outward as a line is shifted into that frequency
by the large velocity gradient.

This behavior follows from the assumption that the conditions and the radiation
field $J_{\nu}$ within the inner zones of the computational grid are
representative for the region below the innermost point.  In cases where a
strong line is present further out this assumption is, however, not justified.
If this is not taken into account by an additional correction, an artificial
emission will build up in the iteration between the moments equation and the
formal ray-by-ray solution.  Eventually this additional emission also affects
the rate equations and the temperature in the inner region because the
respective line transition can pump itself in an unphysical way.

As a first step for an ad hoc correction of this behavior, criteria have to be
established to determine when a correction should be applied. No correction is
needed if the continuum is optically thick or if the local opacity at the first
two radius points is large (e.g., for a strong continuum or if a line
is present at the inner boundary).
Another criterion has to include a comparison of the local opacity to an
average opacity over a reference $\Delta R$-step.  If the average opacity is
higher than the local opacity, a line is likely to be present further out.  The
reference $\Delta R$ is chosen according to a step $\Delta\tau_{\rm c}\ge 3$
for pure continuum opacity (true and Thomson).  The exact conditions for the
correction used in the current implementation are listed in Table~\ref{tab:opi0-corr}.

\begin{table}[b]
  \centering
  \caption{Conditions for the correction of enhanced $J_\nu$ at the inner
    boundary. The reference radius $R_{\rm ref}$ is set to the radius
    of $\tau_{\rm Ross}\approx 2/3$. $R_{\rm ref2}$ refers to the
    radius where $\tau_{\rm c+l}\ge 3$. All $\tau$ values are derived
    radially from the inside outward.}
  \label{tab:opi0-corr}
  \begin{tabular}{@{}lrcr}
    \hline
    Continuum: & $\tau_{\rm c}(R_{\rm ref}) $&$\le$&$ 10$ \\
    Local $\chi$: & $\chi(1)(R(2)-R(1))  $&$\le$&$ 6$   \\
    Compare line to continuum: & $\tau_{\rm c+l}(R_{\rm
      ref2})$/$\tau_{c}(R_{\rm ref2}) $&$\ge$&$ 3$ \\
    Compare local to average $\chi$: & $\tau_{\rm c+l} /
    (\chi(1)(R_{\rm ref} - R(1))) $&$>$&$ 1$\\
    \hline
  \end{tabular}
\end{table}

Secondly, a suitable correction has to be used for each frequency point. We
found that a suitable approach was to use a fixed value $I^{0}_{\nu}=J_{\nu'}$
with $J_{\nu'}$ being $J_{\nu}$ of the previous (redder) uncorrected frequency
point.  To prevent excessively large values for $I^{0}_{\nu}$ in frequency
regions where many subsequent points have to be corrected, an upper cut-off at
$\bnu$ is applied.

\subsubsection{The flux term $H_{\nu}^{0}$}

Instead of the expansion \req{eq:diffapp0} of $S_{\nu}$, we now start
from a general expression for the source function, which explicitly
takes a contribution from Thomson scattering into account
\begin{equation} \label{eq:h0-2}
  S_{\nu}=\left(1-\beta_{\nu}\right)\bnu + \beta J_{\nu}
  \qquad\mr{with}\qquad
  \beta_{\nu}=\frac{\chi_{\nu}^{\rm Th}}{\chi_{\nu}^{\rm Th}+\chi_{\nu}^{\rm true}}.
\end{equation}
From the moments equation in spherical symmetry, \req{eq:momrt3}, one
then gets
\begin{equation} \label{eq:h0-3}
  \sdiff{\left(q_{\nu}f_{\nu}\jti_{\nu}\right)}{\tti_{\nu}}=
  \frac{(1-\beta_{\nu})}{q_{\nu}}\left(\jti_{\nu}-\bti_{\nu}\right)
\end{equation}
(with the Eddington factor $f_{\nu}=J_{\nu}/K_{\nu}$, cf. \req{eq:fnu}) which can be solved analytically by rewriting it to
\begin{equation} \label{eq:h0-5}
    \sdiff{}{\tti_{\nu}}
    \left(q_{\nu}f_{\nu}\left(\jti_{\nu}-\bti_{\nu}\right)\right)=
    \frac{1-\beta_{\nu}}{q_{\nu}^{2}f_{\nu}}
    \left(q_{\nu}f_{\nu}\left(\jti_{\nu}-\bti_{\nu} \right)\right)
\end{equation}
under the assumption that
\begin{equation} \label{eq:h0-4}
  \sdiff{q_{\nu}f_{\nu}\bti_{\nu}}{\tti_{\nu}}\equiv 0\qquad\mr{and}
  \qquad \frac{1-\beta_{\nu}}{q_{\nu}^{2}f_{\nu}}\approx\const.
\end{equation}
The first assumption can be justified by considering only up to first
order terms in an expansion of $q_{\nu}f_{\nu}\bti_{\nu}$ in
$\tti_{\nu}$. The second assumption is not strictly fulfilled, however
in practice a representative mean value
$\langle(1-\beta_{\nu})/(q_{\nu}^{2}f_{\nu})\rangle$ is
used.  The general solution for $J_{\nu}$ in \req{eq:h0-5} is then
derived to
\begin{eqnarray} \label{eq:h0-6}
 q_{\nu}f_{\nu}\jti_{\nu} &=&q_{\nu}f_{\nu}\bti_{\nu} + C_{\nu} \,
  e^{-\sqrt{\left\langle\frac{1-\beta_{\nu}}{q_{\nu}^{2}f_{\nu}}\right\rangle}
    \tti_{\nu}} + C_{\nu}'\nonumber\\
  \Ra\qquad J_{\nu} &=&\bnu + \frac{C_{\nu}}{q_{\nu}f_{\nu}r^{2}} \,
  e^{-\sqrt{\left\langle\frac{1-\beta_{\nu}}{q_{\nu}^{2}f_{\nu}}\right\rangle}
    \tti_{\nu}} + \frac{C_{\nu}'}{q_{\nu}f_{\nu}r^{2}}
\end{eqnarray}
with integration constants $C_{\nu}$ and $C_{\nu}'$ to be determined.
Given the condition that $\tti\ra\infty$, $\jti_{\nu}=\bti_{\nu}$
has to be obtained. It follows that $C_{\nu}'\equiv 0$.

This result can be used to determine the flux term $H_{\nu}^{0}$ from the
moments equation \req{eq:momrt2}, which leads to
\begin{eqnarray} \label{eq:h0-7}
  \hti_{\nu}^{0}&=&\diff{}{\tti_{\nu}}\left(q_{\nu} f_{\nu} r^{2}J_{\nu}\right) \\&=&
   \diff{}{\tti_{\nu}}\left(q_{\nu}f_{\nu}r^{2}B_{\nu}\right) -
   C_{\nu}\sqrt{\left\langle\frac{1-\beta_{\nu}}{q_{\nu}^{2}f_{\nu}}\right\rangle}\,
   e^{-\sqrt{\left\langle\frac{1-\beta_{\nu}}{q_{\nu}^{2}f_{\nu}}\right\rangle}\tti_{\nu}}.
\end{eqnarray}
For the first term one derives the expression
\begin{eqnarray}
  \diff{}{\tti_{\nu}}\left(q_{\nu}f_{\nu}r^{2}B_{\nu}\right)&=&
  \diff{(r^{2}q_{\nu})}{\tti_{\nu}}f_{\nu}B_{\nu} +
  r^{2}q_{\nu}\left(\diff{f_{\nu}}{\tti_{\nu}}B_{\nu} +
  f_{\nu}\diff{B_{\nu}}{\tti_{\nu}}\right)\nonumber\\
  &=& \left\{ -\left(\frac{3f_{\nu}-1}{q_{\nu}\chi_{\nu} r} -
  \diff{f_{\nu}}{\tti_{\nu}}\right)B_{\nu}
  +f_{\nu}\diff{\bnu}{\tti_{\nu}}   \right\}r^{2}q_{\nu} \nonumber\\
  &=& \left\{ -\left(\frac{3f_{\nu}-1}{\chi_{\nu} r} -
  \diff{f_{\nu}}{\tau_{\nu}}\right)B_{\nu}
  +f_{\nu}\diff{\bnu}{\tau_{\nu}}\right\}r^{2} \label{eq:h0-8}
\end{eqnarray}
making use of the definition of the sphericality factor $q_{\nu}$
(\req{eq:sphq}). Note that the last line in \req{eq:h0-8} contains only
derivatives in $\tau_{\nu}$, not $\tti_{\nu}$, because $q_{\nu}$ cancels in
$\dd\tti_{\nu}=-q_{\nu}\chi_{\nu}\,\dd r$.

Next we consider the integration constant $C_{\nu}$ in \req{eq:h0-6}.
This constant can be obtained by considering that in the outer part of
the atmosphere
\begin{equation} \label{eq:h0-9}
  j_{\nu}(\tti_{0})=\left.\frac{H_{\nu}}{J_{\nu}}\right|_{\tti_{0}}\approx 2
    \qquad\mr{for}\quad\tti_{0}(\nu)\ll 1
\end{equation}
holds. It therefore follows that
\begin{equation} \label{eq:h0-10}
  \jti_{\nu}(\tti_{0}(\nu))=j_{\nu}(\tti_{0})
  \left.\diff{q_{\nu}f_{\nu}\jti_{\nu}}{\tti_{\nu}}\right|_{\tti_{0}}.
\end{equation}
Inserting the result of \req{eq:h0-6} into \req{eq:h0-10} gives
\begin{eqnarray} \label{eq:h0-11}
  \lefteqn{\bti_{\nu}(\tti_{0}(\nu)) +
  \frac{C_{\nu}}{(q_{\nu}f_{\nu})_{\tti_{0}}}\,
  e^{-\sqrt{\left\langle\frac{1-\beta_{\nu}}{q_{\nu}^{2}f_{\nu}}\right\rangle}\tti_{0}(\nu)} }\nonumber\\
  &&=j_{\nu}(\tti_{0})
  \left(\left.\diff{\left(q_{\nu}f_{\nu}\bti_{\nu}\right)}{\tti_{\nu}}\right|_{\tti_{0}}
  -
  C_{\nu}\sqrt{\left\langle\frac{1-\beta_{\nu}}{q_{\nu}^{2}f_{\nu}}\right\rangle}
        \, e^{-\sqrt{\left\langle\frac{1-\beta_{\nu}}{q_{\nu}^{2}f_{\nu}}\right\rangle}\tti_{0}(\nu)}\right)
\end{eqnarray}
and hence the expression for $C_{\nu}$ where $\tti_{0}\ll 1$
\begin{equation}\label{eq:h0-12}
  C_{\nu}=\frac{%
    -\bti_{\nu}(\tti_{0}(\nu)) +
    j_{\nu}(\tti_{0}(\nu))\left.
    \diff{\left(q_{\nu}f_{\nu}\bti_{\nu}\right)}{\tti_{\nu}}
    \right|_{\tti_{0}}
  }{%
    \left(\frac{1}{q_{\nu}f_{\nu}}\right)_{\tti_{0}} +
    j_{\nu}(\tti_{0}) \sqrt{\left\langle\frac{1-\beta_{\nu}}{q_{\nu}^{2}f_{\nu}}\right\rangle}
  }\,e^{\sqrt{\left\langle\frac{1-\beta_{\nu}}{q_{\nu}^{2}f_{\nu}}\right\rangle}\tti_{0}(\nu)}.
\end{equation}
In practice, $\tti_{0}$ has been chosen such that, at the corresponding depth
point, the radiation field is not entirely decoupled from matter (e.g.,
$\tti_{0}\approx0.1$).  This is necessary because the temperature in
$\bnu(T(\tti_{0}))$ at this depth point still has to be meaningful to
characterize the radiation field.  Using the result of \req{eq:h0-8}, $C_{\nu}$
can be expressed as
\begin{eqnarray}
  \label{eq:h0-12a}
  C_{\nu} &=& \frac{r(\tti_{0})^{2}
    \left\{j_{\nu}
      \left(
        \diff{f_{\nu}}{\tau_{\nu}}-\frac{3f_{\nu}-1}{\chi_{\nu}r}\right)-1
    \right\}_{\tti_{0}}\bnu(\tti_{0})
    + f_{\nu}(\tti_{0}) \left.\diff{\bnu}{\tau}\right|_{\tti_{0}}
  }{
    \left(\frac{1}{q_{\nu}f_{\nu}}\right)_{\tti_{0}} +  j_{\nu}(\tti_{0})
    \sqrt{\left\langle\frac{1-\beta_{\nu}}{q_{\nu}^{2}f_{\nu}}\right\rangle}} \nonumber\\
  &&{}\times \,e^{\sqrt{\left\langle\frac{1-\beta_{\nu}}{q_{\nu}^{2}f_{\nu}}\right\rangle}
    \tti_{0}(\nu)}.
\end{eqnarray}
As a further approximation we adopt an expansion in $1/\tau_{\nu}$ for $f_{\nu}$
assuming that $f_{\nu}\ra\third$ for $\tau_{\nu}\ra\infty$ to avoid mixing
terms of $\diff{f_{\nu}}{\tau}$ and $f_{\nu}$ in \req{eq:h0-7}:
\begin{equation}
  \label{eq:h0-13}
  f_{\nu}(1/\tau_{\nu}) \approx  \third +
  \diff{f_{\nu}}{(1/\tau_{\nu})} (1/\tau_{\nu}) \quad \Ra \quad
  f_{\nu}(\tau_{\nu})\approx
  \third-\diff{f_{\nu}}{\tau_{\nu}}\tau_{\nu}.
\end{equation}
Introducing \req{eq:h0-13} and \req{eq:h0-8} in \req{eq:h0-7} gives a new
expression for the first term of the inner boundary condition\footnote{
In the equation for the temperature at the inner boundary point
(\req{eq:t1})
accordingly the expression $H^{0}_{\rm new}(\nu)$ has to be used for
$H^{0}_{\nu}$.}
\begin{eqnarray}
  \label{eq:h0-14}
  H^{0}_{\rm new}(\nu)\equiv H^{0}_{\nu} &=& \frac{1}{3}\diff{\bnu}{\tau_{\nu}} +
  \diff{f_{\nu}}{\tau_{\nu}} \left\{
    \left(\frac{3\tau_{\nu}}{\chi_{\nu}R}+1\right)\bnu -
    \tau_{\nu}\diff{\bnu}{\tau_{\nu}}\right\} \nonumber \\
       &&{}-\frac{C_{\nu}}{R^{2}}
       \sqrt{\left\langle\frac{1-\beta_{\nu}}{q_{\nu}^{2}f_{\nu}}\right\rangle}\,
       e^{-\sqrt{\left\langle\frac{1-\beta_{\nu}}{q_{\nu}^{2}f_{\nu}}
           \right\rangle}\tti_{\nu}}.
\end{eqnarray}
This expression resembles the original diffusion flux
$H^{0}_{\rm classic}(\nu)$ plus
additional correction terms that vanish for large optical depths and an
isotropic radiation field.

In the radiative transfer code, a slightly different form was used by solving
\req{eq:h0-13} for $\diff{f}{\tau}$ because the gradient of the Eddington
factor is numerically less accurate than $f_{\nu}$ itself
\begin{equation}
  \diff{f_{\nu}}{\tau_{\nu}} \approx \left(\third-f_{\nu}\right)\tau_{\nu}^{-1}.
\end{equation}
This leads to the alternative form of
\begin{eqnarray}
  \label{eq:h0-15}
   H^{0}_{\rm new'}(\nu) &=&f_{\nu}\diff{\bnu}{\tau_{\nu}} +
  \left(\frac{1}{3}-f_{\nu}\right)
  \left(\frac{1}{\tau_{\nu}}+\frac{3}{\chi_{\nu}R}\right)\bnu
   \nonumber \\
  &&{}-\frac{C_{\nu}}{R^{2}}
  \sqrt{\left\langle\frac{1-\beta_{\nu}}{q_{\nu}^{2}f_{\nu}}\right\rangle}\, e^{-\sqrt{\left\langle\frac{1-\beta_{\nu}}{q_{\nu}^{2}f_{\nu}}
      \right\rangle}\tti_{\nu}}
\end{eqnarray}
In the equation for the temperature at the inner boundary \req{eq:t1}
accordingly the expression $H^{0}_{\rm new'}(\nu)$ has to be used for
$H^{0}_{\nu}$.  The Eddington factor $f_{\nu}$ is calculated from the moments
$J_{\nu}=\int_{0}^1 u_{\nu}(\mu)\,\dd\mu$ and
$K_{\nu}=\int_{0}^1 u_{\nu}(\mu)\,\mu^2\,\dd\mu$ of the radiation field,
obtained from the ray-by-ray solution (which represents an angular-resolved
computation of the intensity, $I_{\nu}(\mu)$).  Note that $f_{\nu}$ is a
quantity not directly imposed by the boundary condition, which only specifies
$I^{+}$ but not $I^{-}$.

\begin{figure}[htb]
  \begin{center}
    \includegraphics[width=8.4cm]{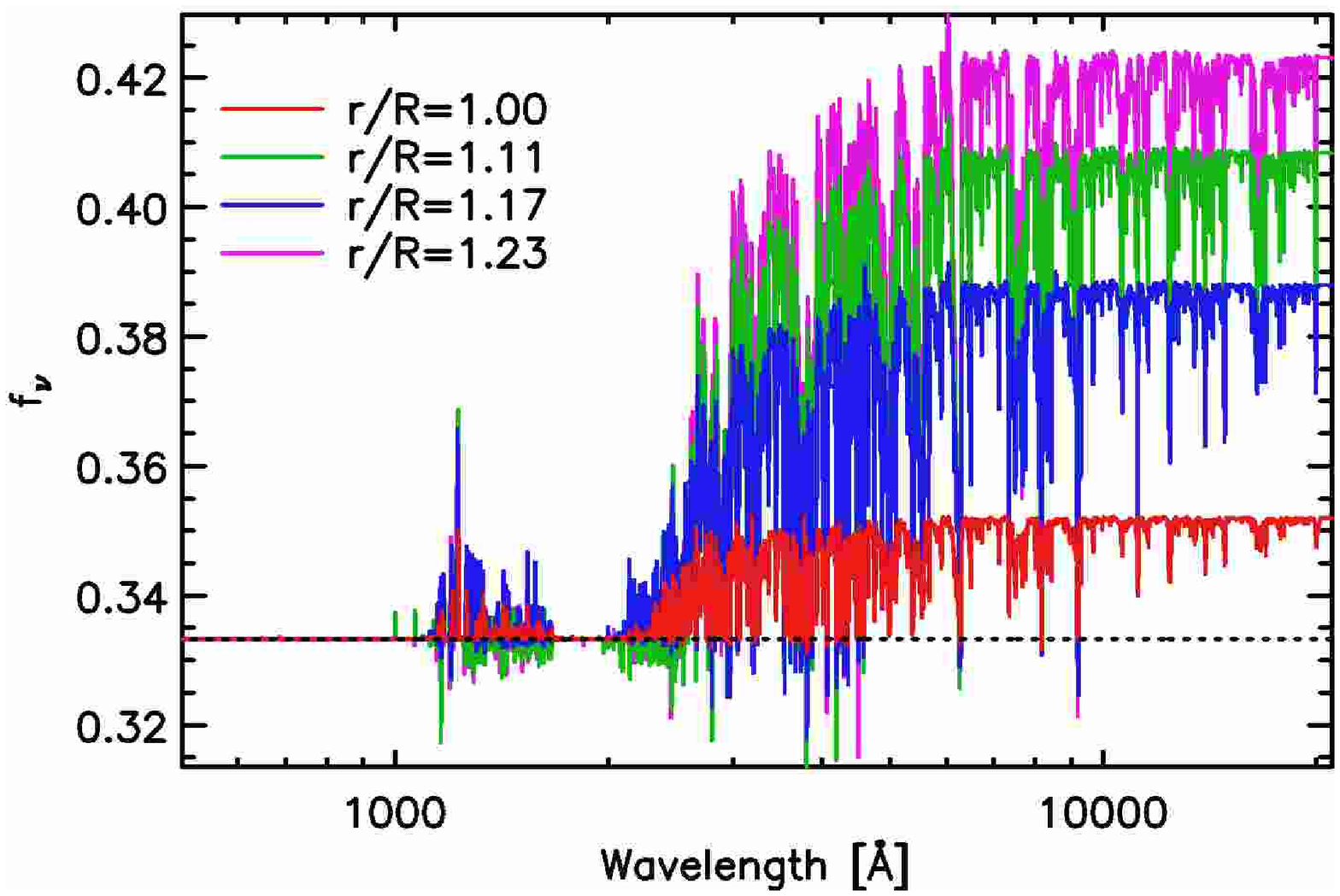}
  \end{center}
  \caption{The Eddington factor $f_{\nu}=J_{\nu}/K_{\nu}$ at the innermost
  radius points as a function of wavelength for the case of a {\snia} model.
  The deviation of $f_{\nu}$ from its  LTE value of $\third$ (indicated by the
  dotted line) allows the new flux term of the inner boundary
  $H^{0}_{\rm new}(\nu)$ to deviate substantially from the LTE value of
  $H^{0}_{\rm classic}(\nu)$. The radii are given in units of the innermost
  radius. } \label{fig:fnu}
\end{figure}

The crucial modification with respect to the classical formulation
\req{eq:diffapp2} of the flux term is achieved by the Eddington factor $f_{\nu}$
that can deviate substantially from its value of $\third$ in the LTE diffusion
limit.  This deviation can be seen in Fig.~\ref{fig:fnu} which shows $f_{\nu}$
as a function of wavelength for the four innermost radial grid points of a
{\snia} model.

\section{Discussion and test calculations}
\label{sec:discussion}

\begin{figure}[htb]
  \centering
  \includegraphics[width=8.4cm]{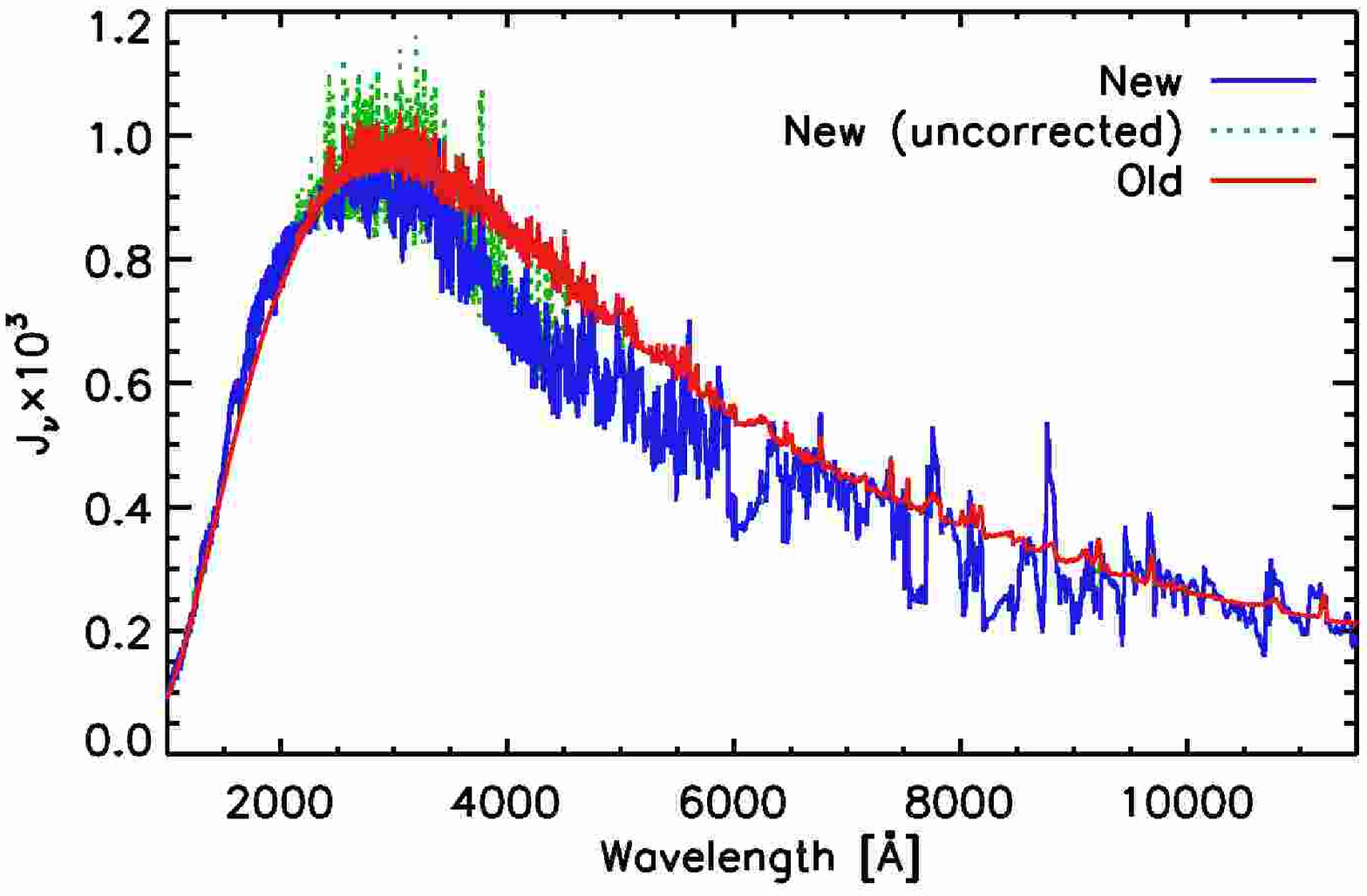}
  \caption{\label{fig:newbc-corr} The radiation field $J_{\nu}$ at the inner
  boundary for different treatments of $I^{+}$. The dotted green line shows the
  radiation field for the uncorrected iterative method, which produces large
  artificial peaks at wavelength points where a line is present further out
  within a small $\tau$-interval. The blue solid line represents the new
  method including the correction for those wavelength points. The red line
  is calculated with the traditional treatment of the inner boundary. It is
  clear that the new method significantly reduces the radiation field in the
  red part of the spectrum. All three models are shown after a few iterations
  before the first temperature update.}
\end{figure}
Fig.~\ref{fig:newbc-corr} shows the radiation field $J_{\nu}$ at the
innermost grid point for the case where the new method is used with and without
correction (indicated by the blue solid line and the green dotted line,
respectively). The third model shown in Fig.~\ref{fig:newbc-corr} uses the
standard boundary condition (red solid line). One can see that the
uncorrected new boundary treatment may create large artificial emission peaks.
These peaks occur at wavelengths where a line is present further out within a
small $\tau$-interval.

Additionally, Fig.~\ref{fig:newbc-corr} clearly shows that the characteristic
of the radiation field is far from Planckian, which causes the standard
diffusion approximation to be inappropriate. Also it can be seen that the new
method produces less radiation in the red and infrared regions compared to the
old boundary treatment.

Looking at the structure of the newly derived flux term \req{eq:h0-15}, we note
that the original flux term of the diffusion approximation \req{eq:diffapp2} is
obtained in the limit of large $\tau$ and $f_{\nu}\ra\third$,  equivalent to
the requirement of a isotropic radiation field.  Under these conditions the
radiation field approaches LTE and the iterated $I_{\nu}^{0}=J_{\nu}$ term will
therefore approach the Planck function $\bnu(T)$. This behavior complies with
the requirement that the original diffusion approximation has to be recovered
for LTE conditions.

\begin{figure}[htb]
  \centering
  \includegraphics[width=8.4cm]{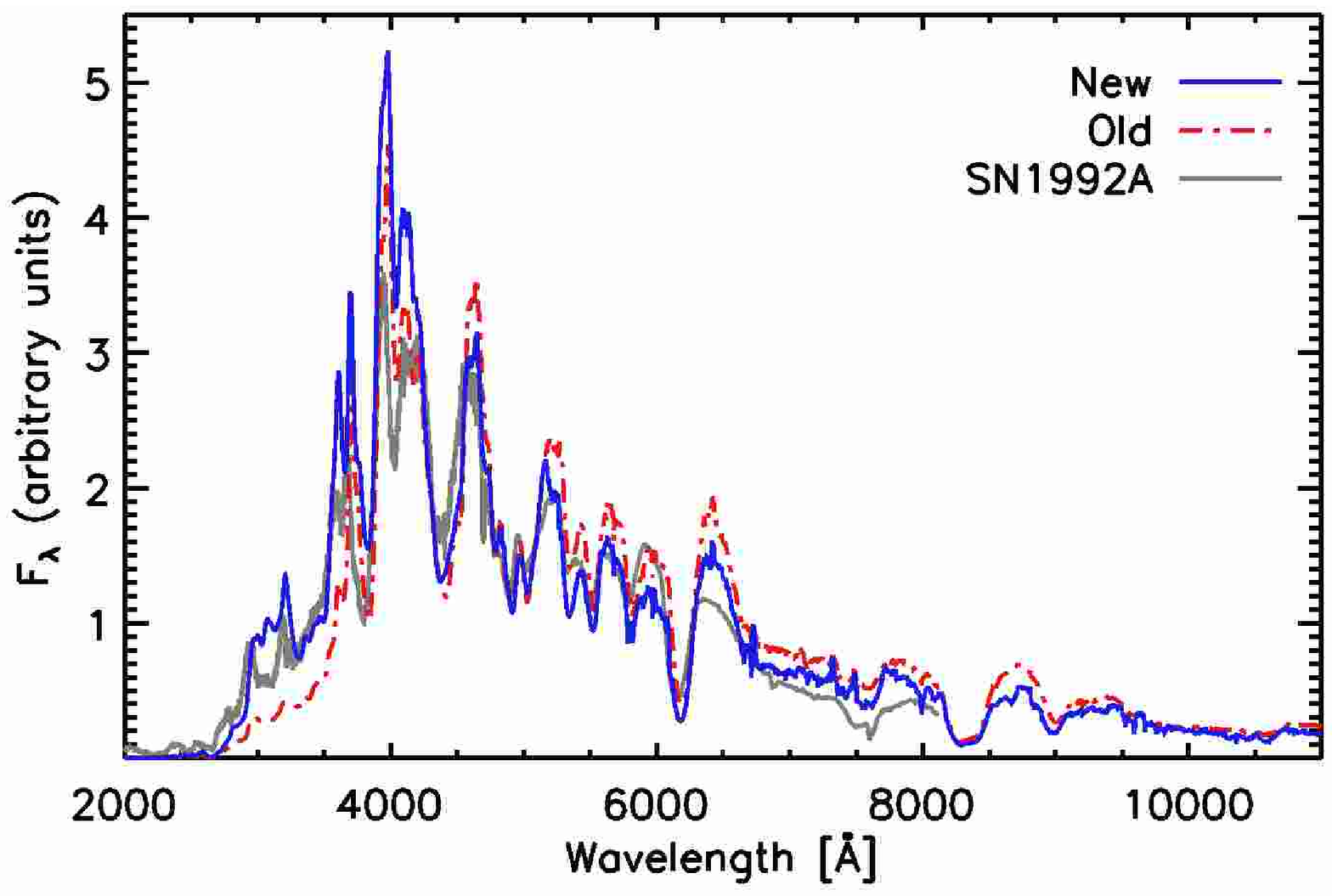}
  \caption{Comparison of two test models using the old and the new
    treatments of the inner boundary. One can see that the flux in the
    red wavelengths is diminished in the model using the new boundary
    treatment. Note that the two models are not strictly comparable
    because the occupation numbers, ionization, and temperature
    structure adjust differently. For comparison the observed spectrum of
    SN~1992A $5\,$d after maximum is shown \citep{kirshner93}.}
  \label{fig:bc-compare}
\end{figure}

Fig.~\ref{fig:bc-compare} shows a comparison of two full test models with the
old and the new treatments of the boundary. One can see that in the red
wavelength region as well as in the peaks of the spectrum, the radiation field
in the model using the new method is slightly diminished compared to the model
using the standard procedure. A direct comparison of the two models, however,
is difficult because the treatment of the boundary condition has a significant
influence on the occupation numbers, the ionization fractions, and the
temperature structure. This influence can be seen especially in the UV part of
the spectrum.  The observed spectrum of SN~1992A $5\,$d after maximum
\citep{kirshner93}, is shown in gray in this figure.  Note that even though
these models are test cases only and have not been tuned to fit the observation
in detail, the overall shape of the observed spectrum is reproduced quite well. 
The model shown here uses the density distribution of the model {\em f1} by
\citet{roepke05} which has been averaged over angles to obtain spherical
symmetry. For the velocity field a homologous expansion law $v\propto r$ is
assumed.  For simplicity, we adopt a `generic', homogeneous composition,
independent of the predictions of the underlying explosion model. Also, the
luminosity was not determined from the {\nifs} content of the explosion model
but set to $L=1\times10^{43}\,${\ergs}.  The entire luminosity is emitted at
the lower boundary implying conservation of the radiation flux through the
ejecta.  Table~\ref{table:model} summarizes the parameters of this model. 
$v_{0}$ is the velocity at the innermost radius of the computational grid.  The
velocity at the `photosphere', $v_{\rm ph}$ (where the photosphere is defined
as the point at which $\tau_{\rm Rosseland}=2/3$), is actually an output
quantity of the model because it depends on the opacities that change with the
occupation numbers over the course of the iterations.  Due to the strong
wavelength-dependence of the opacities, however, this `photospheric velocity'
is not necessarily an observationally meaningful quantity.  We note that the
absorption minimum of the \ion{Si}{ii}$~\lambda6355$ feature in the model
corresponds to a velocity of $\sim\!8\,800\,${\kms} which is significantly
larger than the $v_{\rm ph}$ defined via $\tau_{\rm Rosseland}$. This should be
kept in mind when using $v_{\rm \ion{Si}{ii}}$ to track the photospheric
velocity observationally \citep[e.g.,][]{benetti05,hachinger06}.

\begin{table}
  \centering
  \caption{\label{table:model}Model parameters for the test model shown in
  Fig.~\ref{fig:bc-compare}. The composition is given in fractions by mass.
  The photospheric velocity $v_{\rm ph}$ denotes the velocity where
  $\tau_{\rm Rosseland}=2/3$. $v_{0}$ is the velocity of the actual innermost
  point of the model while $v_{\rm ph}$ is an output of the calculation.}
  \begin{tabular*}{5cm}{l@{\quad}r@{\extracolsep{\fill}}l@{\extracolsep{0pt}\quad}r}
  \hline
  \hline
  C  &  1.1\,\% &   Ti &  0.1\,\%\\
  O  & 11.4\,\% &   Cr &  0.1\,\%\\
  Mg & 10.2\,\% &   Fe &  3.1\,\%\\
  Si & 41.4\,\% &   Co & 11.4\,\%\\
  S  & 13.4\,\% &   Ni &  2.1\,\%\\
  Ca &  5.7\,\% &      & \\[1ex]
  \end{tabular*}
  \begin{tabular*}{5cm}{lr@{ }l}
  $L$          & $1\times10^{43}$   & \ergs \\
  Epoch        & $25\,$             & d  \\
  $v_{0}$      & $4.48\times10^{3}$ & \kms \\
  $v_{\rm ph}$ & $6.65\times10^{3}$ & \kms \\
  \hline
  \end{tabular*}
\end{table}

\section{Conclusions}
\label{sec:conclusion}

We have shown that the physical conditions in the expanding atmospheres of
{\snia} are such that even at early times a complete thermalization of photons
cannot be assumed.  Therefore the common approach using an LTE diffusive
boundary condition at the inner point of the computational grid does not
provide a consistent description of the radiation field and may lead to
non-realistic synthetic spectra.  Observationally this can be seen in the red
and infrared wavelength bands where the spectral slope of typical {\snia}
spectra deviates significantly from a thermal continuum. The assumption of such
a thermal continuum in radiative transfer models generally results in an
overestimate of the radiation flux in those wavelengths. We have developed a
theoretical framework to eliminate some of the restrictions that are imposed by
the assumption of LTE conditions.

With the formalism discussed in this work we are able to self-consistently
derive the isotropic term of the boundary condition. Only the flux term has to
be specified analytically by taking  the physical conditions in the
photospheric region into account.  Of course, removing constraints in the
boundary conditions may cause the system to become numerically less stable.
For this reason the consistency of the solution is much more important than in
the classical case where the explicit specification of the Planck-function
forces the system into LTE-conditions naturally.

Our modifications to the LTE-diffusion approximation have been derived in a
very general way. Therefore, this formalism can also be applied to other
objects (such as Wolf-Rayet stars with very extended atmospheres) where the
physical conditions are such that the assumption of LTE at the photosphere is
not justified.

The comparison of the synthetic spectrum from a {\snia} test model to an
observed {\snia} spectrum shows that the overall shape and prominent features
of observed {\snia} are well reproduced by the model. A more detailed analysis
of observed spectra will be the subject of forthcoming publications.

Late epochs of {\snia} have not been considered here in more detail because the
energy deposition by $\gamma$-photons above the photosphere has not yet been
fully implemented in our code.  The description developed here nevertheless
provides a basis for a reliable implementation of this energy deposition, since
even if a fraction of the radiative energy is created above the photosphere the
remaining radiation has to originate from below the computational grid. 
However, given that the photospheric conditions are such that thermalization
occurs only partially, it is impossible that the energy deposited in the
outer ejecta will be completely thermalized. Instead, excitation and ionization
by fast electrons and $\gamma$-photons above the photosphere may provide a
non-thermal contribution to the spectrum.  This might already be an issue even
for epochs around maximum, as recent three-dimensional explosion models
indicate extensive mixing of {\nifs} into the outer layers of the ejecta.

Furthermore, to derive a model with a luminosity that is consistently
determined from the {\nifs} distribution of an explosion model, the
time-dependent effects of photon trapping in earlier epochs have to be
incorporated into the boundary luminosity \citep{arnett82,hoeflich96,nugent97}.

\begin{acknowledgements}
  This work was supported in part by the Sonderforschungsbereich 375 of the
  \emph{Deut\-sche For\-schungs\-ge\-mein\-schaft, DFG\/}, the European Union's
  Human Potential Programme {\em ``Gamma-Ray Bursts: An Enigma and a Tool,''}
  under contract HPRN-CT-2002-00294, and the National Science Foundation under
  Grant No.  PHY99-07949.  D.~N.~S. thanks the Kavli Institute for Theoretical
  Physics at the University of California, Santa Barbara for its hospitality
  during the program on ``The Supernova Gamma-Ray Burst Connection.''  We also
  want to thank our colleague Wolfgang Hillebrandt for helpful discussions and
  his support during the course of this work.
\end{acknowledgements}

\bibliographystyle{aa}
\bibliography{jabref}

\end{document}